\documentclass[prc,letterpaper,twocolumn,showpacs,showkeys,lengthcheck,tightenlines,
               nofootinbib,preprintnumbers,superscriptaddress]{revtex4-1} 

\pdfpageattr {/Group << /S /Transparency /I true /CS /DeviceRGB>>}
\usepackage{newtxtext,newtxmath}

\usepackage{titlesec}
\usepackage{dcolumn}
\usepackage{bm}
\usepackage{amsmath,amssymb,amsfonts}

\usepackage{graphicx}

\usepackage{color}
\usepackage{bbold}

\usepackage{slashed}
\usepackage[svgnames]{xcolor}
\usepackage[english]{babel}
\usepackage{blindtext}
\usepackage{microtype}
\usepackage{tikz}

\usepackage[colorlinks=true,linkcolor=blue,urlcolor=blue,citecolor=blue]{hyperref}
\usepackage{ifpdf}
\usepackage[utf8]{inputenc}

\def\a{\alpha}
\def\b{\beta}

\def\g{\gamma}

\def\ve{\varepsilon}

\def\l{\lambda}

\def\G{\Gamma}
\def\L{\Lambda}

\def\hs{\hspace}

\def\ol{\overline}
\def\no{\nonumber}

\def\lf{\left}
\def\rg{\right}
\def\la{\langle}
\def\ra{\rangle}

\newcommand{\vect}[1]{\boldsymbol{#1}}
\newcommand{\ph}[1]{\phantom{#1}}

\newcommand{\sh}[1]{\slashed{#1}}

\font\bb=bbmss10 scaled 1200
\def\ident{\mbox{\bb 1}}

\graphicspath{{.}{figures/}} 

\titlespacing{\section}{4pt}{10pt plus 4pt minus 2pt}{8pt plus 2pt minus 2pt}
\titlespacing{\subsection}{0pt}{12pt plus 4pt minus 2pt}{8pt plus 2pt minus 2pt}
\begin{document}

\title{TMDs of Spin-one Targets: Formalism and Covariant Calculations}

\author{Yu Ninomiya}
\email[]{yuu.n.2038@gmail.com}
\affiliation{Department of Physics, School of Science, Tokai University,
             4-1-1 Kitakaname, Hiratsuka-shi, Kanagawa 259-1292, Japan}
             
\author{Wolfgang Bentz}
\email[Corresponding author:~]{bentz@keyaki.cc.u-tokai.ac.jp}
\affiliation{Department of Physics, School of Science, Tokai University,
             4-1-1 Kitakaname, Hiratsuka-shi, Kanagawa 259-1292, Japan}
\affiliation{Radiation Laboratory, Nishina Center, RIKEN, Wako, Saitama 351-0198,
Japan}

\author{Ian~C. Cloët}
\affiliation{Physics Division, Argonne National Laboratory, Argonne, Illinois 60439, USA}

\begin{abstract}
We present a covariant formulation and model calculations of the leading-twist time-reversal even transverse momentum-dependent quark distribution functions (TMDs) for a spin-one target. Emphasis is placed on a description of these 3-dimensional distribution functions which is independent of any constraints on the spin quantization axis. We apply our covariant spin description to all nine leading-twist time-reversal even $\rho$ meson TMDs in the framework provided by the Nambu--Jona-Lasinio model, incorporating important aspects of quark confinement via the infrared cut-off in the proper-time regularization scheme.  In particular, the behavior of the 3-dimensional TMDs in a tensor polarized spin-one hadron are illustrated. Sum rules and positivity constraints are discussed in detail. Our results do {\it not} exhibit the familiar Gaussian behavior in the transverse momentum, and other results of interest include the finding that the tensor polarized TMDs -- associated with spin-one hadrons -- are very sensitive to quark orbital angular momentum, and that the TMDs associated with the quark operator $\g^+\vect{\g}_T\g_5$ would vanish were it not for dynamical chiral symmetry breaking. In addition, we find that 44\% of the $\rho$ meson's spin is carried by the orbital angular momentum of the quarks, and that the magnitude of the tensor polarized quark distribution function is about 30\% of the unpolarized quark distribution. A qualitative comparison between our results for the tensor structure of a quark-antiquark bound state is made to existing experimental and theoretical results for the two-nucleon (deuteron) bound state.
\end{abstract} 


\maketitle
%===============================================================================
%===============================================================================
\section{INTRODUCTION\label{sec:intro}}
Quarks and gluons are confined inside hadrons through the strong interaction, therefore, high energy scattering experiments are required to probe the quark and gluon structure of hadrons and hadronization processes. This makes parton distribution functions (PDFs) and fragmentation functions (FFs) crucial ingredients in hadron and particle physics~\cite{Collins:1981uw,Jaffe:1996zw,Collins:2011zzd}. Recently, transverse momentum-dependent quark distribution functions (TMDs) and fragmentation functions (TMD FFs) have been of significant focus  both experimentally and theoretically~\cite{Barone:2001sp,Accardi:2012qut,Metz:2016swz}. At leading-twist the internal transverse momentum-dependent quark structure of spin-half hadrons, such as the  nucleon, is expressed in terms of six time-reversal even ($T$-even) and two time-reversal odd ($T$-odd) TMDs, and after  integrating over the transverse momenta of quarks there remain three PDFs: the unpolarized,  helicity and transversity PDFs.  However, for spin-one hadrons, such as vector mesons or the  deuteron, the spin degrees of freedom require three additional leading-twist $T$-even TMDs, resulting in one  additional PDF~\cite{Hoodbhoy:1988am,Hino:1999qi,Bacchetta:2000jk}.\footnote{For spin-one hadrons there are also seven additional $T$-odd TMDs~\cite{Bacchetta:2000jk}.% which we do not consider herein.
} 

In the pioneering work of Hoodbhoy, Jaffe and Manohar~\cite{Hoodbhoy:1988am}, this new PDF was called $b_1^q(x)$, where $x$ is the Bjorken scaling variable and $q$ the quark flavor. It represents the difference of unpolarized quark distributions in a longitudinally polarized spin-one hadron with spin projection $\l=0$ and $\l=\pm 1$. In the notation of Ref.~\cite{Hoodbhoy:1988am}\footnote{In the Bjorken limit the corresponding structure function is  $b_1(x) = \sum_q \, e_q^2\lf[b_1^q(x) + b_1^{\bar{q}}(x)\rg]$, where $e_q$ are the quark electric charges.}
\begin{align}
b_1^q(x) &= \frac{1}{4} \lf[2\,q^{(\l=0)}(x) - q^{(\l=1)}(x) - q^{(\l=-1)}(x)\rg], 
\label{definition-b1}
\end{align}
where $q^{(\l)}(x) = q^{(\l)}_{\uparrow} + q^{(\l)}_{\downarrow}$  are the unpolarized quark distribution functions.  It has been pointed out in Ref.~\cite{Hoodbhoy:1988am}, and in subsequent investigations~\cite{Close:1990zw,Umnikov:1996qv}, that $b_1^q(x)$ is sensitive to the orbital motion of the constituents, and may also be sensitive to QCD exotica~\cite{Miller:1989nc,Liuti:2014dda,Miller:2013hla}. An important sum rule for the $b_1^q$ quark distribution is derived in Refs.~\cite{Efremov:1981vs,Close:1990zw} and reads
\begin{align}
\int^1_0 dx\lf[b_1^q(x) - b_1^{\bar{q}}(x)\rg] = 0,
\label{ck}
\end{align}
which simply means that the valence quark number does not depend on the hadron's spin state.  It has been shown in Ref.~\cite{Umnikov:1996qv} that Lorentz covariance is important to satisfy the $b_1^q$ sum rule, even though the constituents may be nonrelativistic. 

There have been several experimental and theoretical studies of the $b_1$ structure function for spin-one targets. For the deuteron, $b_1(x)$ was first measured by the HERMES Collaboration~\cite{Airapetian:2005cb}, and a  phenomenological parametrization of the data was performed in Ref.~\cite{Kumano:2010vz}, which found that sea quarks significantly contribute to $b_1(x)$ in the low $x$ region.  On the theoretical side, the $b_1$ structure function of the deuteron has been investigated using, e.g., convolution approaches~\cite{Hoodbhoy:1988am,Khan:1991qk,Miller:2013hla,Cosyn:2017fbo} and the covariant Bethe-Salpeter equation~\cite{Umnikov:1996qv}. Because of its importance, more precise experimental data for the deuteron $b_1$ structure function will be taken at Jefferson Lab as part of the 12\,GeV program~\cite{Slifer:2014bda,E1213011}.  For vector mesons like the $\rho$, there exists no experimental data, however several theoretical studies have been performed, examples include using $\rho$-meson light-cone wave functions~\cite{Mankiewicz:1988dk}, and lattice QCD to determine the first few moments~\cite{Best:1997qp}.

The generalization of the PDF defined in Eq.~\eqref{definition-b1}, together with its corresponding quark fragmentation function, to include transverse momentum-dependence was performed in Refs.~\cite{Hino:1999qi,Bacchetta:2000jk}.  In the latter paper, the spin density matrix formalism was used to derive important positivity constraints  for the TMDs and PDFs of spin-one hadrons, by extending the methods used for the spin-half case~\cite{Bacchetta:2001rb}.  To the best of our knowledge, however, it is not yet clear how to systematically apply those theoretical formulations to model calculations using effective theories of QCD. This is an important motivation for the present study.

In this paper we present a covariant formulation for the TMDs of a spin-one target which is independent of any constraints on the spin quantization axis. This formalism is then used in a covariant model calculation for all nine $T$-even TMDs of the $\rho^+$ meson. In deriving these results we first establish the connection between the polarization 4-vectors, used in standard Feynman diagram calculations, and the spin 4-vectors and tensors which are used to define the various spin-one TMDs. Special care is taken to present a clear and workable definition of the three tensor polarized TMDs: $f_{LL}(x,\vect{k}_T^2)$, $f_{LT}(x,\vect{k}_T^2)$, $f_{TT} (x,\vect{k}_T^2)$, where $\vect{k}_T$ is the quark transverse momentum and we use the same notation as Ref.~\cite{Bacchetta:2001rb}.\footnote{We omit the commonly used subscript ``$1$'' on all TMDs because we only consider leading-twist.} 
The relation between these TMDs and the $b_1^q$ quark distribution of Eq.~\eqref{definition-b1} will be derived, and the positivity constraints on the spin-one TMDs and PDFs will be discussed in detail. 

The formalism presented here can be straightforwardly used to determine the TMDs of a spin-one hadron in any framework used to model QCD, and we will choose the Nambu--Jona-Lasinio (NJL) model~\cite{Nambu:1961tp,Nambu:1961fr} as an effective quark theory for QCD. A key feature of the NJL model is that chiral symmetry is dynamically broken, and it has also been extended to incorporate important aspects of quark confinement through an infrared cut-off in the proper-time regularization scheme~\cite{Ebert:1996vx,Hellstern:1997nv,Bentz:2001vc}. Our results for the nine $T$-even TMDs and four PDFs of the $\rho^+$ meson respect all requirements of Lorentz covariance at the quark level. The NJL model has also been successfully used to describe PDFs and electromagnetic form factors of mesons and baryons, e.g., see Refs.~\cite{Mineo:1999eq,Cloet:2005rt,Cloet:2007em,Ninomiya:2014kja,Cloet:2014rja,Carrillo-Serrano:2016igi,Carrillo-Serrano:2015uca}. 

Experimental data for $\rho$-meson TMDs will not be available for the foreseeable future, however, it is nevertheless important to study them in various non-perturbative approaches to QCD. In addition, spin-one TMDs are a useful passage to the spin-one TMD FFs, which may be measurable in the production of vector mesons~\cite{Ji:1993vw}. In fact, by using crossing symmetry the TMD FFs for an elementary quark fragmentation process to vector mesons can be obtained from the TMDs. We remark further on this point in Appendix~\ref{ex-form}. These elementary TMD FFs may then be used as input to integral equations which describe multi-fragmentation processes~\cite{Bentz:2016rav}. Further, the description of the nucleon TMDs and TMD FFs, which are extensively studied experimentally and theoretically~\cite{Metz:2016swz}, often requires the inclusion of spin-one (vector and axial-vector) diquarks. One may therefore use the results of our present investigation as part of a description of the nucleon TMDs and TMD FFs.      

This paper is organized as follows: Sect.~\ref{form-Nino} presents the formalism for the leading-twist time-reversal even TMDs of a spin-one target, and also provides the relations between the polarization vectors, and the spin vectors and tensors which are needed in actual calculations. Sect.~\ref{NJL} briefly introduces the NJL model, presents the covariant bound state description for the $\rho$-meson, and describes the calculation of the $\rho$-meson TMDs. Sect.~\ref{results} provides a detailed discussion of our results and a summary is given in Sect.~\ref{Conclusion}.

%===============================================================================
%===============================================================================
\section{QUARK TMDS AND PDFS FOR SPIN-ONE TARGETS\label{form-Nino}}
This section first presents several useful relations for polarization and spin 4-vectors and tensors, and then discusses the operator definitions of TMDs and PDFs for spin-one targets. 

\subsection{Polarization and Spin 4-vectors}
The polarization 4-vectors of a spin-one particle with 4-momentum $p$ and mass $m_h$ are given by
\begin{align}
\ve_{(\l)}^{\mu}(p) = \lf( \frac{\vect{p} \cdot \vect{\ve}_{(\l)}}{m_h}, \,
\vect{\ve}_{(\l)} + \frac{\vect{p} \lf(\vect{p} \cdot \vect{\ve}_{(\l)}\rg)}{m_h\,(E_p +m_h)} \rg),
\label{emu}
\end{align}
where $E_p = \sqrt{\vect{p}^2 + m_h^2}$. The polarization 3-vectors $\vect{\ve}_{(\l)} $ are eigenvectors of $\vect{\Sigma} \cdot \vect{S}$ with eigenvalues $\l=\pm 1, 0$, where $\vect{\Sigma}$ are the $3 \times 3$ spin matrices in the adjoint representation with components $\lf(\Sigma_k\rg)_{ij} = -i \ve_{kij}$, and the unit vector $\vect{S}$ represents the spin quantization axis, which can be chosen arbitrarily. The orientation of the polarization 3-vectors is chosen so that $\vect{\ve}_{(0)} = \vect{S}$, $\vect{\ve}_{(\pm 1)} \cdot \vect{S} = 0$, and 
$-i \lf(\vect{\ve}_{(\l)}^* \times \vect{\ve}_{(\l)} \rg) = \l\,\vect{\ve}_{(0)}$.\footnote{If one chooses $\vect{S}$ along the $z$-axis, the polarization 3-vectors assume the standard forms  $\vect{e}_{(+1)} = -\frac{1}{\sqrt{2}} \lf(1, i, 0\rg)$, $\vect{e}_{(-1)} = \frac{1}{\sqrt{2}} \lf(1, -i, 0\rg)$,  and $\vect{e}_{(0)} = \lf(0, 0, 1\rg)$.  For a general direction of $\vect{S}$ characterized by the polar angles $(\theta, \phi)$, the polarization 3-vectors can be obtained by a rotation expressed in the notation of Euler angles by $\vect{\ve}_{(\l)} = e^{-i \phi \Sigma_3} \, e^{-i \theta \Sigma_2} \, e^{i \phi \Sigma_3}\, \vect{e}_{(\l)}$.} By construction the polarization 4-vectors satisfy the following relations:
\begin{align}
\label{sum}
\sum_{\l} \ve_{(\l)}^{\mu*} (p)\, \ve_{(\l)}^{\nu}(p)
= - g^{\mu \nu} + \frac{p^{\mu} p^{\nu}}{m_h^2},
\end{align}
\vspace*{-1.3em}
\begin{align} 
\label{r4a}
\ve_{(\l)}^{*{\mu}}(p) \, \ve_{(\l) \mu}(p) &= -1, & \ve_{(\l)}^{\mu}(p) \, p_{\mu} &= 0, \\[0.5em]
\label{r4b}
\ve_{(1)}^{\mu *}(p) &= -\ve_{(-1)}^{\mu}(p),  & \ve_{(0)}^{\mu *}(p) &=  \ve_{(0)}^{\mu}(p). 
\end{align}
We identify the spin 4-vector of a spin-one particle by
\begin{align}
S^{\mu}(p) = \lf( \frac{\vect{p} \cdot \vect{S}}{m_h}, \,
\vect{S} + \frac{\vect{p} \lf(\vect{p} \cdot \vect{S} \rg)}{m_h (E_p +m_h)} \rg),
\label{s}
\end{align}
which is identical to $\ve_{(0)}^{\mu}(p)$.\footnote{More precisely,  the spin 4-vector for a state with $\l=\pm 1, 0$ is equal to $\l\, S^{\mu}(p)$ with $S^{\mu}(p)$ given by Eq.~\eqref{s}. Note that the state $\l=0$ has zero vector polarization but non-zero tensor polarization.} The following relation between the spin and polarization 4-vectors is then easy to confirm~\cite{Jaffe:1996zw,Jaffe:1988up}:
\begin{align}
\l\,S^{\mu}(p) = \frac{i}{m_h}\, \ve^{\mu \nu \alpha \beta} p_{\nu} \ \ve_{(\l)\alpha}^*(p) \ \ve_{(\l)\beta}(p).
\label{jaf}
\end{align}
TMDs are defined to be independent of the polarization state of the target, we therefore assume pure spin states characterized by the spin projection $\lambda$ on an arbitrary spin quantization axis $\vect{S}$. This also provides for a clearer physical interpretation of the TMDs.
For the Feynman diagram calculations that give the TMDs it is then convenient to express the spin density matrix $\ve^{\mu *}_{(\l)}(p)\,\ve^{\nu}_{(\l)}(p)$, for pure spin states with some fixed $\l$, by the spin and momentum 4-vectors. To derive such a result it is convenient to decompose the spin density matrix into symmetric and antisymmetric Lorentz tensors. The antisymmetric piece can then be expressed by the spin 4-vector using Eq.~\eqref{jaf}, and for the symmetric part one can use the completeness relation of Eq.~\eqref{sum} to express it as a second rank spin tensor~\cite{Jaffe:1996zw,Jaffe:1988up}. The final result is the following identity:
\begin{align}
&\ve_{(\l)}^{\mu *} (p)\ \ve_{(\l)}^{\nu} (p) 
= \frac{1}{3} \lf( -g^{\mu \nu} + \frac{p^{\mu}p^{\nu}}{m_h^2} \rg)
-\frac{i\l}{2\,m_h}\, \ve^{\mu\nu\alpha\beta}p_{\alpha}S_{\beta} (p) \no \\
&\hs{12mm}
-\frac{3\l^2-2}{2}
\lf[S^{\mu}(p)S^{\nu}(p) - \frac{1}{3} \lf( -g^{\mu \nu} +\frac{p^{\mu}p^{\nu}}{m_h^2} \rg) \rg].
\label{formula}
\end{align}

In this study we choose the 3-momentum of the target in the $z$-direction and denote the Cartesian components of the unit vector $\vect{S}$ by $\vect{S} = \lf(\vect{S}_T, S_L\rg)$, where $\vect{S}_T=(S_T^1, S_T^2)$ is normal to the hadron momentum and $S_L=S^3$ is the $z$-component. In terms of $\vect{S}_T$ and $S_L$ the spin 4-vector given in Eq.~\eqref{s} is expressed as
\begin{align}
S^{\mu}(p) = \lf( \frac{p^3}{m_h} \, S_L ,\, \vect{S}_T , \, \frac{E_p}{m_h} S_L \rg),
\label{sp}
\end{align}  
where for any given direction $\vect{S}$, the particle can have the three possible spin projections $\l=\pm 1, 0$ onto this direction.  Therefore, longitudinal polarization means that $\vect{S}_T=0$ and $|S_L|=1$, and there are three spin projections $\l=\pm 1, 0$ along the direction of the momentum, whereas transverse polarization means that $S_L=0$ and $|\vect{S}_T|=1$, and there are  three spin projections  $\l=\pm 1, 0$ along the direction $\vect{S}_T$ perpendicular to the momentum. We emphasize that $S_L$ and $\l$ are in general different quantities.

%===============================================================================
%===============================================================================
\subsection{Definition of Spin-one TMDs and PDFs}
\begin{figure}
\centering\includegraphics[width=\columnwidth]{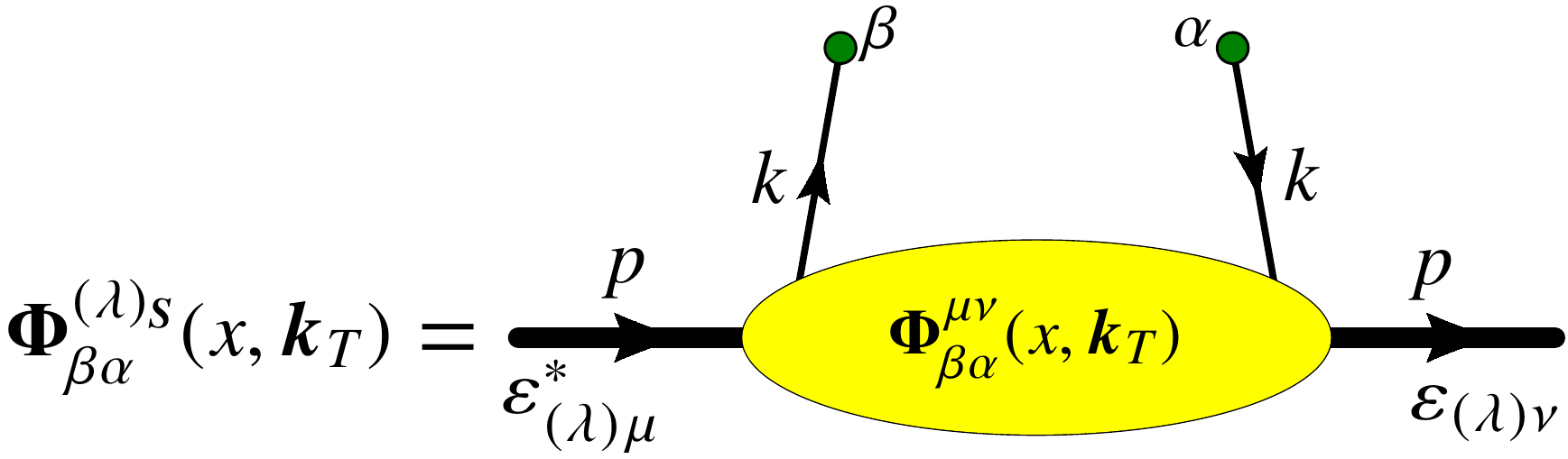}
\caption{(Color online) Graphical representation of the correlator of Eq.~\eqref{phi1}. The dots labeled by $\alpha$, $\beta$ indicate
the Dirac indices of the quark field operators, the line labeled by the momentum $k$ represents
the active quark, and the thick line labeled by the momentum $p$ and the polarizations $\ve^{\mu *}$ and
$\ve^{\nu}$ represents its parent spin-one target.
The shaded oval represents the spectator states.}
\label{quark-correlation-function}
\end{figure}

The TMDs of a spin-one target are defined with respect to the following transverse momentum-dependent quark correlation function:
\begin{align}
\Phi_{\beta \alpha}^{(\l)_{\vect{S}}}(x, \vect{k}_T) &= 
%\frac{1}{p^+} \int \frac{dk^+ dk^-}{(2\pi)^4} \delta\lf(x - \frac{k^+}{p^+}\rg)  \no\\
\int \frac{dk^+ dk^-}{(2\pi)^4} \delta\lf(k^+ - x\,p^+\rg)  \no\\
& \hs{6mm}
\times \int d^4 z \ e^{i k\cdot z}  \  
{}_{\vect{S}}\langle p,\l | \overline{\psi}_{\alpha}(0) \, \psi_{\beta}(z) | p, \l \rangle_{\vect{S}}, \no\\
&= \int \frac{dz^- \, d^2 \vect{z}_T}{(2\pi)^3} \, 
e^{i\lf(xp^+\,z^- - \,\vect{k}_T \cdot\, \vect{z}_T\rg)}   \no\\
&\hs*{17mm}
\times \ {}_{\vect{S}}\langle p,\l | \overline{\psi}_{\alpha}(0) 
\psi_{\beta}(z^-,\vect{z}_T) | p, \l \rangle_{\vect{S}}, \no \\[0.2em]
&\equiv \ve^*_{(\l) \mu}(p) \ \Phi_{\beta\alpha}^{\mu\nu}(x,\vect{k}_T) \ \ve_{(\l) \nu}(p),
\label{phi1}
\end{align}
which we illustrate in Fig.~\ref{quark-correlation-function}.\footnote{This TMD definition ignores the additional complications that arise from explicit gluon degrees of freedom and the gauge link.} In Eq.~\eqref{phi1} $\psi$ represents the flavor SU(2) quark field operator, $\a$ and $\beta$ are Dirac indices, and $a^{\pm} = \tfrac{1}{\sqrt{2}}\lf(a^0 \pm a^3\rg)$, $\vect{a}_T=(a^1, a^2)$ are the light-cone $\pm$ and transverse components of a 4-vector $a^{\mu}$.     The 3-momentum of the target ($\vect{p}$) is assumed in the $z$-direction, and the quark's momentum components normal to this direction are denoted by $\vect{k}_T$. The state $|p, \l \rangle_{\vect{S}}$ indicates that the projection of the target's spin on the direction $\vect{S}$ is equal to $\l=\pm 1, 0$, and is normalized according to the covariant normalization $_{\vect{S}}\langle p, \l'|p, \l \rangle_{\vect{S}} = 2\,p^+ V \delta_{\l' \l}$, where $V$ is the quantization volume.  In the last step of Eq.~\eqref{phi1} we expressed the quark correlation matrix as a contraction of  the polarization-independent Lorentz tensor matrix $\Phi_{\beta\alpha}^{\mu\nu}$ and the product of polarization 4-vectors for the external target lines. 

%=====================================================================
%=====================================================================

The matrix $\Phi^{(\l)_{\vect{S}}}(x, \vect{k}_T)$ of Eq.~\eqref{phi1} is decomposed into Dirac $\g$-matrices as usual by~\cite{Barone:2001sp}
\begin{align} \label{exp}
\Phi_{\beta \alpha}^{(\l)_{\vect{S}}} = \frac{1}{2} \lf[s + v^{\mu}\, \g_{\mu} 
+ a^{\mu}\,\g_5 \g_{\mu} + \frac{i}{2}\,t^{\mu \nu}\, \g_5\,\sigma_{\mu \nu}\rg]_{\beta \alpha}.
\end{align}
%
% At leading-twist the following coefficient functions contribute:
% %
% \begin{align}
% v^{+} (x,\vect{k}_T) &= \frac{1}{2} \, {\rm Tr}_D \lf[\g^+ \, \Phi^{(\l)_{\vect{S}}}(x, \vect{k}_T)\rg], \no \\
% &\equiv \langle \g^+ \rangle^{(\l)}_{\vect{S}} (x,\vect{k}_T), \no \\  
% &\equiv  \ve^*_{(\l) \mu}(p) \  \lf<\g^{+} \rg>^{\mu\nu}\! (x,\vect{k}_T) \
% \ve_{(\l) \nu} (p),
% \label{vp}  \\[0.6em]
% %
% %
% a^+ (x,\vect{k}_T) &= \frac{1}{2} \, {\rm Tr}_D \lf[\g^+ \g_5 \, \Phi^{(\l)_{\vect{S}}}(x, \vect{k}_T)\rg], \no\\
% &\equiv \langle \g^+ \g_5 \rangle^{(\l)}_{\vect{S}} (x,\vect{k}_T), \no \\
% &\equiv  \ve^*_{(\l) \mu}(p)  \ \lf<\g^{+} \g_{5}\rg>^{\mu\nu}\! (x,\vect{k}_T) \
% \ve_{(\l) \nu} (p), \label{ap} \\[0.6em]
% %
% %
% t^{+i} (x,\vect{k}_T) &= \frac{1}{2} \, {\rm Tr}_D \lf[-i \sigma^{+i} \g_5\,\Phi^{(\l)_{\vect{S}}}(x, \vect{k}_T)\rg], \no \\
% &\equiv \langle \g^+ \g^i \g_5 \rangle^{(\l)}_{\vect{S}} (x,\vect{k}_T), \no \\
% &= \ve^*_{(\l) \mu} (p) \  \lf<\g^{+} \g^{i} \g_{5}\rg>^{\mu\nu}\! (x,\vect{k}_T) \
% \ve_{(\l) \nu} (p),  \label{tp} 
% \end{align}
%
The coefficient functions that contribute at leading-twist are $v^+$, $a^+$ and $t^{+i}$, which we represent respectively by
\begin{align}
\label{vp}
\langle \g^+ \rangle^{(\l)}_{\vect{S}} (x,\vect{k}_T) 
&= \frac{1}{2} \, {\rm Tr}_D \lf[\g^+ \, \Phi^{(\l)_{\vect{S}}}(x, \vect{k}_T)\rg], \no \\
&\hs*{-7mm}
\equiv  \ve^*_{(\l) \mu}(p) \  \lf<\g^{+} \rg>^{\mu\nu}\! (x,\vect{k}_T) \ \ve_{(\l) \nu} (p),  \\[0.6em]
\label{ap}
\langle \g^+ \g_5 \rangle^{(\l)}_{\vect{S}} (x,\vect{k}_T) 
&= \frac{1}{2} \, {\rm Tr}_D \lf[\g^+ \g_5 \, \Phi^{(\l)_{\vect{S}}}(x, \vect{k}_T)\rg], \no\\
&\hs*{-7mm}
\equiv  \ve^*_{(\l) \mu}(p)  \ \lf<\g^{+} \g_{5}\rg>^{\mu\nu}\! (x,\vect{k}_T) \ \ve_{(\l) \nu} (p),  \\[0.6em]
\label{tp} 
\langle \g^+ \g^i \g_5 \rangle^{(\l)}_{\vect{S}} (x,\vect{k}_T)
&= \frac{1}{2} \, {\rm Tr}_D \lf[-i \sigma^{+i} \g_5\,\Phi^{(\l)_{\vect{S}}}(x, \vect{k}_T)\rg], \no \\
&\hs*{-7mm}
\equiv \ve^*_{(\l) \mu} (p) \  \lf<\g^{+} \g^{i} \g_{5}\rg>^{\mu\nu}\! (x,\vect{k}_T) \ \ve_{(\l) \nu} (p),  
\end{align}
where $i=1,2$ denotes the transverse vector components. From the structure of Eq.~\eqref{formula}, and rotational invariance in the transverse plane, we can parametrize these coefficient functions by the spin-one hadron TMDs in the same way as
Ref.~\cite{Bacchetta:2001rb}:\footnote{In our calculations of Sect.~\ref{NJL} only the $T$-even TMDs are non-zero, therefore we do not display the $T$-odd TMDs in the parametrizations of Eqs.~\eqref{form1}--\eqref{form3}.}
\begingroup
\allowdisplaybreaks
\begin{align}
&\langle \gamma^+ \rangle_{\vect{S}}^{(\lambda)} (x, \vect{k}_T^2) \equiv f(x, \vect{k}_T^2) + S_{LL} \,
f_{LL}(x, \vect{k}_T^2) \no \allowdisplaybreaks[0] \\
&\hs{2mm} 
+ \frac{\vect{S}_{LT} \cdot \vect{k}_T}{m_h}\, f_{LT}(x, \vect{k}_T^2) 
+ \frac{\vect{k}_T \cdot \vect{S}_{TT} \cdot \vect{k}_T}{m_h^2} \, f_{TT}(x, \vect{k}_T^2) \,,
\label{form1}   \\
&\hs*{0mm}\lf<\g^{+} \g_{5}\rg>^{(\l)}_{\vect{S}}\! (x,\vect{k}_T)  \equiv  \l \lf [S_L\,g_L(x,\vect{k}_T^2) 
+ \frac{\vect{k}_T\cdot \vect{S}_T}{m_h}\, g_T(x,\vect{k}_T^2)  \rg], 
\label{form2}  \\
&\lf<\g^+\g^i\g_{5}\rg>^{(\l)}_{\vect{S}}\! (x,\vect{k}_T)  \equiv \l 
\Biggl[S_T^i\,h(x,\vect{k}_T^2) 
+ S_L\,\frac{k_T^i}{m_h}\,h_L^\perp(x,\vect{k}_T^2) 
\no  \\
&\hs{15mm}
+
\frac{1}{2\,m_h^2}
\lf(2\,k_T^i\,\vect{k}_T \cdot \vect{S}_T - S_T^i\,\vect{k}_T^2\rg) h_T^{\perp}(x,\vect{k}_T^2) \Biggr], 
%+ \frac{\vect{k}_T^2}{2m_h^2} \left(-S_T^i + 2 \frac{k_T^i \left(\vect{k}_T \cdot \vect{S}_T\right)}
%{\vect{k}_T^2}\right) \, h_T^{\perp}(x,\vect{k}_T^2) \Biggr], 
%
%
\label{form3}
\end{align}
\endgroup
where we have introduced the following quantities with implicit $\vect{S}$ and  $\lambda$
dependence:
\begin{align}
S_{LL} &= \lf(3 \lambda^2 - 2\rg) \left(\tfrac{1}{6} - \tfrac{1}{2}\,S_L^2\right), 
\label{sll}  \\
S_{LT}^i &=  \lf(3 \lambda^2 - 2\rg) S_L\, S_T^i,
\label{slt}  \\
S_{TT}^{ij} &= \lf(3 \lambda^2 - 2\rg) \lf(S_T^i\,S_T^j - \tfrac{1}{2}\,\vect{S}_T^2 \, \delta^{ij} \rg).
\label{stt} 
\end{align}
The last three terms on the right-hand side of Eq.~\eqref{form1} define the tensor polarized TMDs, which are specific to hadrons with spin $J \geqslant 1$. In calculations these tensor polarized TMDs arise by contracting the expression obtained from the Feynman diagrams for $\langle \g^+ \rangle ^{\mu \nu}(x, \vect{k}_T)$ with the tensor part of Eq.~\eqref{formula}. 

A partonic (probabilistic) interpretation of the quark TMDs can be obtained by using the light-cone quantization formalism~\cite{Jaffe:1991ra}. Following the same steps as in Sect.~II of Ref.~\cite{Bentz:2016rav},  we arrive at the relation
\begin{align}
\hs*{-1mm}
\langle \G^{a} \rangle_{\vect{S}}^{(\l)}(x, \vect{k}_T) \, dx \, d^2 \vect{k}_T =
\frac{_{\vect{S}}\langle p,\l | b^{\dagger}(\vect{k}) \sigma^{a} \, b(\vect{k}) | p, \l\rangle_{\vect{S}}}
{_{\vect{S}}\langle p,\l | p, \l \rangle_{\vect{S}}},
\label{part}
\end{align}
where $\vect{k} = \lf(x p^+, \vect{k}_T \rg)$ and we have arranged the four Dirac matrices into the 4-vector:
\begin{align} 
\G \equiv \lf\{\g^+,\,\g^+\g^1\g_5,\,\g^+\g^2\g_5,\,\g^+\g_5\rg\},  
\label{gamma}
\end{align} 
and also defined $\sigma^{a} = \lf(1, \vect{\sigma}\rg)$ with $\vect{\sigma}$ the usual Pauli spin matrices. For the quark creation and annihilation operators, $b^{\dagger}$ and $b$, we have used the notation  $b^{\dagger}(\vect{k})\, \sigma^{a} \, b(\vect{k}) \equiv \sum_{s' s} b^{\dagger}_{s'}(\vect{k})\,  \lf(\sigma^{a}\rg)_{s' s} b_s(\vect{k})$. It then follows from Eq.~\eqref{part} that for $a=0$ the quantity on the left-hand side is the number density of quarks with longitudinal momentum fraction $x$ and transverse momentum $\vect{k}_T$, while for $a=1,2,$ or $3$ it is the number density of quarks with $(x, \vect{k}_T)$ and spin in the $x, y,$ or $z$ direction, minus the same number density but for opposite quark spin direction.

All distributions defined so far refer to the TMDs of a quark. The TMDs of an antiquark are defined in terms of the correlation function given in Eq.~\eqref{phi1}, except with the quark field $\psi$ replaced by the charge conjugated field $\psi_c = C\,\ol{\psi}{}^T = i \g_2\,\psi^*$, where $C = i \g_2 \g_0$ is the charge conjugation matrix. Then, by using translational invariance and the relation $C^{-1} \G^T C = \{-\g^+, -\g^+ \g^i \g_5, \g^+ \g_5 \}$, one can show that this antiquark correlation function for some  $(x,\,\vect{k}_T)$ with $0 \leqslant x \leqslant 1$, is equal to the correlation function of Eq.~\eqref{phi1}, but evaluated at $(-x, -\vect{k}_T)$ and attaching a minus sign for the operators $\g^+$ and $\g^+ \g^i \g_5$.

Integrating our results for $\langle \G \rangle^{(\l)}_{\vect{S}} (x,\vect{k}_T)$ [Eqs.~\eqref{form1}--\eqref{form3}] over $\vect{k}_T$ gives the four PDFs for a spin-one target:
\begingroup
\allowdisplaybreaks
\begin{align}
\label{pdf1}
\langle \g^{+} \rangle^{(\l)}_{\vect{S}} (x)  
&\equiv   f(x)  + S_{LL} \, f_{LL}(x),  \\[0.1ex]
\label{pdf2}
\lf<\g^{+} \g_{5}\rg>^{(\l)}_{\vect{S}} (x) 
&\equiv \l\, S_{L} \, g (x), \\[0.1ex]
\label{pdf3}
\lf<\g^{+}\g^{i} \g_{5}\rg>^{(\l)}_{\vect{S}} (x) 
&\equiv \l \, S_T^{i} \, h (x).
\end{align}
\endgroup
%
% \begin{align}
% \langle \g^{+} \rangle^{(\l)}_{\vect{S}} (x)  
% &= \ve^*_{(\l) \mu}(p) \, \langle \g^{+} \rangle^{\mu \nu}(x) \, \ve_{(\l) \nu} (p) \no \\ 
% % 
% &\equiv   f(x)  + S_{LL} \, f_{LL}(x), \label{pdf1} \\[0.8ex]
% %
% %
% \lf<\g^{+} \g_{5}\rg>^{(\l)}_{\vect{S}} (x) 
% &= \ve^*_{(\l) \mu}(p) \, \langle \g^{+} \g_5 \rangle^{\mu \nu}(x) \, \ve_{(\l) \nu}(p) \no \\
% &\equiv \l\, S_{L} \, g (x), \label{pdf2} \\[0.8ex]
% %
% %
% \lf<\g^{+}\g^{i} \g_{5}\rg>^{(\l)}_{\vect{S}} (x) 
% &= \ve^*_{(\l) \mu}(p) \, \langle \g^{+} \g^i \g_5 \rangle^{\mu \nu}(x) \, \ve_{(\l) \nu}(p) \no \\
% &\equiv \l \, S_T^{i} \, h (x). \label{pdf3} 
% \end{align}
%
The relations between the TMDs and PDFs are given by
\begingroup
\allowdisplaybreaks
\begin{align}
\label{fll}
\lf[f,f_{LL},g,h\rg](x) &= \int d^2 \vect{k}_T \lf[f,f_{LL},g_L,h\rg](x,\vect{k}_T^2).
%
%f(x) &= \int d^2 \vect{k}_T \, f(x,\vect{k}_T^2), \label{f} \\
%f_{LL}(x) &= \int d^2 \vect{k}_T \, f_{LL}(x, \vect{k}_T^2), \label{fll} \\ 
%g(x) &= \int d^2 \vect{k}_T \, g_{L}(x,\vect{k}_T^2), \label{g} \\
%h(x) &= \int d^2 \vect{k}_T  \, h(x,\vect{k}_T^2)\,. \label{h}  
\end{align}
\endgroup
Therefore, for spin-one hadrons there are nine leading-twist TMDs and four PDFs, if we restrict ourselves to the $T$-even distributions. Note that the tensor polarized PDF $f_{LL}(x)$, defined in Eq.~\eqref{fll}, is related to $b_1^q(x)$ given in  Eq.~\eqref{definition-b1} by
\begin{align}
b_1^q(x) = \frac{1}{2} \, f_{LL}^q(x),
\label{bt}
\end{align}
where we have made explicit the quark flavor index. This result is easily seen by noting that in the helicity basis $[S_L=1, \,\vect{S}_T=0]$ our quantity $\langle \g^+ \rangle^{(\l)}_{\vect{S}}(x)$ of Eq.~\eqref{pdf1} becomes equal to $f(x) - f_{LL}(x)\,(3 \l^2 -2)/3$. Taking the combination of hadron helicities $\l$ indicated in Eq.~\eqref{definition-b1}, and using the probabilistic interpretation of $\langle \g^+ \rangle_{\vect{S}}^{(\l)}(x)$ expressed by Eq.~\eqref{part}, it is easy to confirm Eq.~\eqref{bt}.

To conclude this section we discuss the positivity constraints among the spin-one hadron TMDs and PDFs. From the probabilistic interpretation, given by Eq.~\eqref{part}, of the quantities $\lf<\G \rg>_{\vect{S}}^{(\l)}(x,\vect{k}_T^2)$ of Eqs.~\eqref{form1}--\eqref{form3}, we obtain the inequalities:
\begin{align}
\label{eq:positivityf}
\lf<\g^{+} \rg>_{\vect{S}}^{(\l)}(x,\vect{k}_T^2) &\geqslant 0, \\
\label{positiveg}
\lf<\g^{+} \rg>_{\vect{S}}^{(\l)}(x,\vect{k}_T^2) &\geqslant
\lf|\lf <\g^{+}\g_{5}\rg >_{\vect{S}}^{(\l)}(x,\vect{k}_T^2) \rg|,\\
\label{positiveh}
\lf<\g^{+} \rg>_{\vect{S}}^{(\l)}(x,\vect{k}_T^2) &\geqslant
\lf|\lf <\g^{+}\g^{i}\g_{5}\rg >_{\vect{S}} ^{(\l)}(x,\vect{k}_T^2) \rg|.
\end{align}
These relations must be valid for any spin quantization axis $\vect{S}$ and spin projection $\l$. These inequalities lead, e.g., to the following {\it naive} positivity constraints for the spin-one TMDs:\footnote{By {\it naive} positivity constraints we refer to constraints derived from Eqs.~\eqref{eq:positivityf}--\eqref{positiveh}, as opposed to the more rigorous positivity constraints obtained from the antiquark--hadron forward scattering matrix (see Appendix~\ref{app:relations}). }
\begin{align}
\label{eq:naive1}
\hs*{-1mm}
f(x,\vect{k}_T^2) &\geqslant  0, \\
\label{eq:naive2}
\hs*{-1mm}
-\frac{3}{2} f(x,\vect{k}_T^2) &\leqslant f_{LL}(x,\vect{k}_T^2) \leqslant 3\,f(x,\vect{k}_T^2), \\
\label{eq:naive3}
\hs*{-1mm}
\lf|g_L(x,\vect{k}_T^2)\rg| &\leqslant  f(x,\vect{k}_T^2) - \frac{1}{3}\,f_{LL}(x,\vect{k}_T^2) \leqslant \frac{3}{2}\,f(x,\vect{k}_T^2), \\
\label{eq:naive4}
\hs*{-1mm}
\lf|h(x,\vect{k}_T^2)\rg| &\leqslant  f(x,\vect{k}_T^2) + \frac{1}{6}\,f_{LL}(x,\vect{k}_T^2) \leqslant \frac{3}{2}\,f(x,\vect{k}_T^2),
%
%
%\lf|h_L^\perp(x,\vect{k}_T^2)\rg| &\frac{\lf|\vect{k}_T\rg|}{2\,m_h} \leqslant  f(x,\vect{k}_T^2) - \frac{1}{3}\,\theta(x,\vect{k}_T^2) \leqslant \frac{3}{2}\,f(x,\vect{k}_T^2), \\
%
%\label{eq:naive6}
%\lf|g_T(x,\vect{k}_T^2)\rg| &\frac{\lf|\vect{k}_T\rg|}{2\,m_h} \leqslant  f(x,\vect{k}_T^2) + \frac{1}{6}\,\theta(x,\vect{k}_T^2) \leqslant \frac{3}{2}\,f(x,\vect{k}_T^2).
\end{align}
where Eqs.~\eqref{eq:naive1}--\eqref{eq:naive3} are derived from Eqs.~\eqref{eq:positivityf} and \eqref{positiveg} by considering a spin-one hadron with longitudinal polarization $[\vect{S}_T=0,\,|S_L|=1]$, and Eq.~\eqref{eq:naive4} is derived from Eq.~\eqref{positiveh} by considering the case of transverse polarization $[|\vect{S}_T|=1,\,S_L=0]$.  A more general discussion of positivity constraints among the TMDs, based on the antiquark--hadron forward scattering matrix, can be found in Appendix~\ref{app:relations} and Ref.~\cite{Bacchetta:2001rb}. In these analyses it is shown that positivity of the eigenvalues of the antiquark--hadron forward scattering matrix leads to six {\it sufficient} positivity conditions on the spin-one TMDs. These sufficient conditions are rather complicated and difficult to use in practice, however numerous additional {\it necessary} conditions can be obtained for the spin-one TMDs by considering the positivity of the principal minors of the antiquark--hadron forward scattering matrix, and Eqs.~\eqref{eq:positivityf}--\eqref{positiveh} as explained above. Integrating Eqs.~\eqref{eq:naive1}--\eqref{eq:naive4} over $\vect{k}_T$ gives the following {\it naive} positivity conditions on the PDFs:
\begingroup
\allowdisplaybreaks
\begin{align}
&{\rm I}. & f(x) &\geqslant  0, 
\label{bound_PDF-1} \\
&{\rm II}. &  -\frac{3}{2} f(x) &\leqslant f_{LL}(x) \leqslant 3\,f(x), 
\label{bound_PDF-2}  \\
&{\rm III}. & |g(x)| &\leqslant f(x) - \frac{1}{3}\,f_{LL}(x) \leqslant \frac{3}{2}\,f(x)  ,
\label{bound_PDF-3} \\
&{\rm IV.} & |h(x)| &\leqslant  f(x) + \frac{1}{6}\,f_{LL}(x) \leqslant \frac{3}{2}\,f(x) .
\label{bound_PDF-4}
\end{align}
\endgroup
Analysis of the antiquark--hadron forward scattering matrix for the PDFs gives three {\it sufficient} conditions (see Appendix~\ref{app:relations}), which can be expressed as Eqs.~\eqref{bound_PDF-1}--\eqref{bound_PDF-3} and the generalization of the Soffer bound~\cite{Soffer:1994ww} for spin-one targets [see Eq.~\eqref{soff1}],
%
%\begin{align}
%2\,h(x)^2 \leqslant \lf(f(x) + \frac{2}{3}\,\theta(x)\rg) 
%\lf(f(x) + g(x) - \frac{1}{3}\,\theta(x)\rg),
%\label{eq:soff1} 
%\end{align}
%
which is a more stringent version of Eq.~\eqref{bound_PDF-4}.

%===============================================================================
%===============================================================================
\section{ NAMBU--JONA-LASINIO MODEL AND RHO TMDS\label{NJL}} 
This section presents the method used to calculate the $\rho$-meson TMDs and PDFs. We use the NJL model, which is a covariant quantum field theory that exhibits dynamical chiral symmetry breaking and, via the introduction of an infrared cut-off in the proper-time regularization scheme, incorporates important aspects of the quark confinement. Further details are given in Appendix~\ref{ex-form}. 

%===============================================================================
%===============================================================================
\subsection{Rho Meson as a Relativistic Bound State}
The two flavor NJL model Lagrangian is given by
\begin{align}
\mathcal{L}_{NJL} &= \overline{\psi}(i\sh{\partial}-m )\psi +G_{\pi} \lf[ (\overline{\psi} \psi)^2 - 
(\overline{\psi}\g_5\vect{\tau} \psi)^2 \rg] \no\\
&\hs*{-4mm}
- G_{\omega} (\overline{\psi}\g^{\mu}\psi)^2 
- G_{\rho} \lf[ (\overline{\psi}\g^\mu\,\vect{\tau} \psi)^2 
+ (\overline{\psi}\g^\mu\g_5\vect{\tau} \psi)^2  \rg],
\end{align}
where $m = \text{diag}[m_u,m_d]$ is the current quark mass matrix, $\vect{\tau}$ are the Pauli matrices for isospin SU(2), and $G_{\pi}, G_{\omega}$ and $G_{\rho}$ are the 4-fermion coupling constants in the $\pi$, $\omega$, and $\rho$ meson channels. The dynamically generated dressed quark mass $M$ is obtained via the gap equation:
\begin{align}
M = m + 48 i \, G_{\pi} \, M \,   \int  \frac{d^4 k}{\lf(2\pi\rg)^4}\ \frac{1}{k^2 - M^2 + i \ve}, 
\label{gap}
\end{align}
which has a preferred solution in the Nambu-Goldstone phase for $G_{\pi}$ larger than a critical coupling.

\begin{figure}[tbp]
\centering\includegraphics[width=\columnwidth]{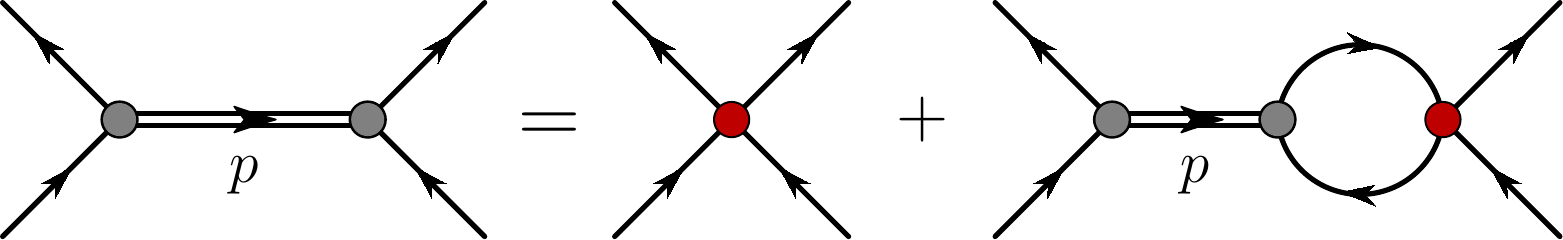}
\caption{(Color online) The $\bar{q}q$ Bethe-Salpeter equation in the random phase approximation. The double-line represents the $\bar{q}q$ $t$-matrix, the single line is the dressed quark propagator and the dark shaded circle is the $\bar{q}q$ interaction kernel.}
\label{BS}
\end{figure}

A description of the $\rho$-meson as a relativistic $\bar{q}q$ bound state is provided by the Bethe-Salpeter equation, which is illustrated in Fig.~\ref{BS}, and reads
\begin{align}
&\mathcal{T}_{\g\delta,\alpha\beta} (p) = \mathcal{K}_{\g\delta,\alpha\beta}  \no \\
&\hs{3mm}
+ \int \frac{d^4 k}{(2\pi)^4} \mathcal{K}_{\g\delta,\l \l'}\, S(k)_{\ve' \l'}\, S(p+k)_{\l \ve}\,
\mathcal{T}_{\ve \ve',\alpha\beta}(p). 
\label{bs}
\end{align}
The solution to this equation provides the $\bar{q}q$ $t$-matrix in the $\rho$-meson channel, where $\mathcal{K} = -2i\,G_{\rho}\,(\g^{\mu} \tau_i) (\g_{\mu} \tau_{i}) $ is the NJL interaction kernel, and  $S(k) = \lf[\sh{k} - M + i\ve \rg]^{-1}$ represents the dressed quark propagator. The solution of Eq.~\eqref{bs} reads
\begin{align}
\mathcal{T}_{\g\delta,\alpha\beta} (p) &= [\g_{\mu}\tau_{i}]_{\g\delta}\ \frac{-2i\,G_{\rho}}{1 + 2\,G_{\rho}\,\Pi_{\rho}(p^2)} \no\\
&\hs{11mm}
\lf(g^{\mu\nu} + 2\,G_{\rho}\,\Pi_{\rho}(p^2) \frac{p^{\mu}p^{\nu}}{p^2}\rg)
[\g_{\nu}\tau_{i}]_{\alpha\beta},
\label{eq:rhotmatrix}
\end{align}
where we have defined the bubble diagram $\Pi_{\rho}(p^2)$ as 
\begin{align}
&\Pi_{\rho}(p^2) \lf( g^{\mu\nu}-\frac{p^{\mu}p^{\nu}}{p^2} \rg) \no \\
&\hs*{16mm}
= 6i \int \frac{d^4 k}{(2\pi)^4}\ {\rm Tr}_D\!\lf[\g^{\mu}\,S(k)\,\g^{\nu}\,S(p+k)\rg].
\label{bub}
\end{align}
The physical mass of the rho, $m_{\rho}$, is obtained by identifying the pole in the $\rho$-meson $t$-matrix of Eq.~\eqref{eq:rhotmatrix}:
\begin{align}
1 + 2\,G_{\rho}\,\Pi_{\rho}(p^2 = m_{\rho}^2) = 0.
\end{align}
The residue of the $t$-matrix at the pole $p^2 = m_{\rho}^2$ defines the meson--quark-antiquark vertex functions, $\G^{(\l),i}_{\a\b}$. Expanding Eq.~\eqref{eq:rhotmatrix} about $p^2 = m_{\rho}^2$ gives
\begin{align}
\mathcal{T}_{\g\delta,\alpha\beta} (p) &\longrightarrow \lf[{Z_\rho} \tau_{i} \g^{\mu}\rg]_{\g\delta} 
\frac{i\lf( g_{\mu\nu} - \frac{p_{\mu}p_{\nu}}{p^2}   \rg)}{p^2-m_{\rho}^2+i\ve} 
\lf[Z_\rho\,\tau_i\,\g^{\nu}\rg]_{\alpha\beta} \no\\
& \equiv  \sum_{\l=0,\pm 1} \overline{\G}^{(\l),i}_{\g\delta}\ \frac{i}{p^2 - m_\rho^2 + i\ve}\ \G^{(\l),i}_{\alpha\beta}.
\end{align}
Therefore the $\rho$ vertex and conjugate vertex functions read
\begin{align}
\overline{\G}^{(\l),i}_{\g\delta}\  \G^{(\l),i}_{\alpha\beta} = 
\lf[i\,Z_\rho\,\tau_i\,\g^{\mu}\,\ve^*_{(\l) \mu}\rg]_{\g\delta}
\lf[i\,Z_\rho\,\tau_i\,\g^{\nu}\,\ve_{(\l) \nu}\rg]_{\alpha\beta},
\label{eq:bsvertex}
\end{align}
where the $\rho$--quark-antiquark coupling constant is given by
\begin{align}
Z_\rho^{-2} \equiv  - \Pi'_{\rho}(p^2=m_{\rho}^2) \,,
\label{grho}
\end{align}
and the prime denotes differentiation with respect to $p^2$.

%=====================================================================
%=====================================================================

%===============================================================================
%===============================================================================
\subsection{Calculation of the Rho Meson TMDs and PDFs \label{sec:rhoTMDs}}
\begin{figure}[tbp]
\centering\includegraphics[width=\columnwidth]{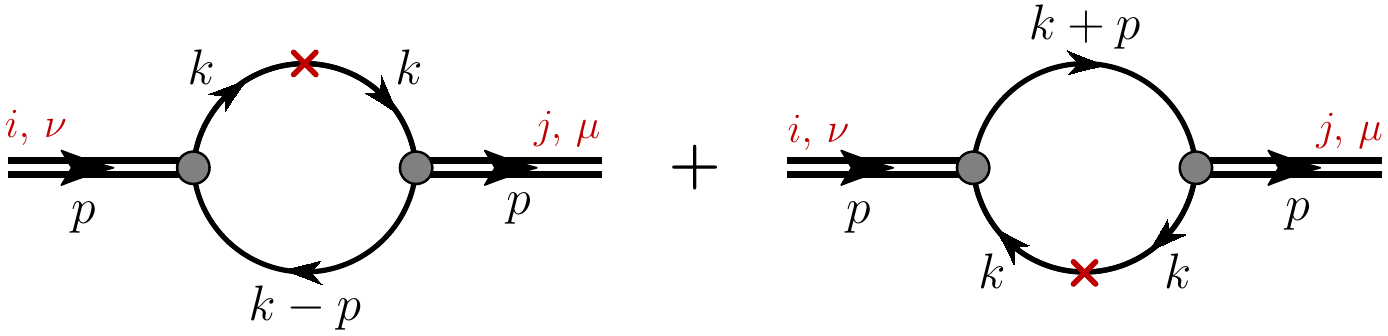}
\caption{(Color online) Feynman diagrams for the $\rho$-meson TMDs in the NJL model. The shaded circles represent the $\rho$-meson Bethe-Salpeter vertex functions of Eq.~\eqref{eq:bsvertex} and the solid lines the dressed quark propagator.  The operator insertion has the form $\G \,\delta \lf(x-\frac{k^+}{p^+} \rg)$, with $\G$ given in Eq.~\eqref{gamma}. The left diagram gives the valence $u$ quark TMDs and the right diagram the valence $\bar{d}$ antiquark TMDs in the $\rho^+$ meson.}
\label{NJL-feynman}
\end{figure}

From causality it follows that the product of field operators on the light-cone ($z^+=0$) in the operator definition of Eq.~\eqref{phi1} can be replaced by the time-ordered product, and therefore we can represent the Lorentz tensors $\lf<\G\rg>^{\mu\nu}\!(x, \vect{k}_T)$ which appear in Eqs.~\eqref{vp}--\eqref{tp} by Feynman diagrams. It is therefore straightforward to use the NJL model to calculate $\lf<\G\rg>^{\mu\nu}\!(x, \vect{k}_T)$, and this calculation corresponds to the one-loop Feynman diagrams given in Fig.~\ref{NJL-feynman}. The first diagram of Fig. \ref{NJL-feynman} has support for $0\leqslant x \leqslant 1$ and directly leads to the valence quark TMDs. Whereas, the second diagram  has support for $-1 \leqslant x \leqslant 0$ and, by using the prescriptions explained in Sect. II.B, leads to the valence antiquark TMDs. A straightforward application of those prescriptions to the present NJL model calculation shows  that all valence antiquark TMDs are exactly the same functions as the valence quark TMDs. In the following we will therefore refer to the valence quark TMDs (first diagram of Fig. \ref{NJL-feynman}) only.

For the case of the $\rho^+$ meson, application of the Feynman rules to the first diagram
of Fig. \ref{NJL-feynman} gives for the valence $u$ quark: 
\begin{align}
\lf<\G \rg>^{\mu\nu} (x,\vect{k}_T)&= -\frac{3i\,Z_\rho^2}{p^+} \int \frac{dk^+dk^-}{(2\pi)^4}
\ \delta \lf(x-\frac{k^+}{p^+} \rg) \no \\
&\hs*{11mm}
{\rm Tr}_D \lf[\g^{\mu} S(k) \,\G \, S(k) \g^{\nu} S(k-p) \rg].
\label{NJL-form}
\end{align}
The methods of covariant integration for expectation values of local operators can be used within the proper-time regularization scheme by first applying a Mellin transformation to Eq.~\eqref{NJL-form}, which gives the $n$-th moment of $\lf<\G\rg>^{\mu\nu}$ as   
\begin{align}
\langle \G \rangle_n^{\mu\nu} (\vect{k}_T) &\equiv \int^1_0 dx \, x^{n-1}\, \lf<\G \rg>^{\mu\nu}\!(x,\vect{k}_T),
\label{NJL-tensor}
\end{align}
where $n=1,\,2,\ldots$ is an integer. These moments can then be determined by applying standard methods, such as, introducing Feynman parameters and shifting the loop momentum to give a quadratic form in the denominators. Because of the simplicity of the NJL model, the TMDs defined by Eq.~\eqref{NJL-form} can then be extracted analytically from  results for the moments~\cite{Mineo:2003vc}. This method preserves Lorentz covariance, which is particularly important for satisfying the sum rules. Contracting the results for $\lf<\G \rg>^{\mu\nu}\!(x, \vect{k}_T)$ with the Lorentz tensor given by Eq.~\eqref{formula}, and parametrizing the expressions as defined by Eqs.~\eqref{form1}--\eqref{form3}, gives the $\rho^+$ meson TMDs. Explicit results for $\lf<\G \rg>^{\mu\nu}\!(x, \vect{k}_T)$ and the TMDs are given in Appendix~\ref{ex-form}.

%=====================================================================
%=====================================================================

An important feature of our NJL model calculation is that all sum rules are explicitly satisfied. For example, the sum rules for the first moment of the spin-independent $f(x)$ and tensor polarized $f_{LL}(x)$ PDFs follow from the following Ward identity for $\langle \g^+ \rangle^{\mu \nu}(x)$ [see Eq.~\eqref{pdf1}]:
\begin{align}
\int_0^1 dx \, \langle \g^+ \rangle^{\mu \nu}(x)
&= Z_\rho^2 \, \frac{1}{2\,p^+}  \frac{\partial}{\partial p_+} \, \Pi_{\rho}(p^2) \lf(g^{\mu \nu} -
\frac{p^{\mu} p^{\nu}}{p^2} \rg)  \no   \\
&\hs*{-15mm}
= Z_\rho^2 \, \lf(g^{\mu \nu} - \frac{p^{\mu} p^{\nu}}{p^2} \rg) \,  \Pi_{\rho}'(p^2) \no \\
&\hs*{-10mm}
+ Z_\rho^2 \, \lf(- \frac{g^{\mu +} p^{\nu} + g^{\nu +} p^{\mu}}{2\,p^+} + \frac{p^{\mu} p^{\nu}}{p^2} \rg)
\frac{\Pi_{\rho}(p^2)}{p^2},
\label{ward}
\end{align}
where $p^2 = m_{\rho}^2$. 
If we contract with the polarization $4$-vectors of the hadron, and use Eq.~\eqref{r4a} we obtain
\begin{align}
\int_0^1 dx \, \langle \g^+ \rangle^{(\l)}_{\vect{S}}(x)
= - Z_\rho^2 \, \Pi'_{\rho}(p^2) = 1,
\label{norm}
\end{align}
where Eq.~\eqref{grho} is used to obtain unity. Comparison with the right-hand side of Eq.~\eqref{pdf1} then gives the sum rules
\begin{align}
\int_0^1 \, dx \, f(x) &= 1, &
\int_0^1 \, dx \, f_{LL}(x) &= 0.   \label{sumft}
\end{align}
The sum rule for $f(x)$ simply means that in our model we have one valence $u$ quark in the $\rho^+$ meson, and the same result is obtained for the number sum rule for the valence antiquark.\footnote{One should note that in general for a meson only the (zero) baryon number, which is the difference of quark and antiquark numbers, is conserved. A separate number sum rule for the valence quark and antiquark holds only in valence-like quark models, such as our present NJL model.} The sum rule for $f_{LL}(x)$ corresponds to the $b_1$ sum rule of Eq.~\eqref{ck} in our valence quark model. We stress that satisfying the Ward identity given in Eq.~\eqref{ward} is crucial to respecting the sum rules.
In particular, the Lorentz tensor structure of  the fully integrated $\langle \g^+ \rangle^{\mu \nu}$ must be the same as the transverse Lorentz tensor structure of the polarization (bubble) diagram of Eq.~\eqref{bub}. Because our calculations are completely covariant, these constraints are always satisfied, as the explicit expressions in Appendix~\ref{ex-form} demonstrate. 

Our results for the functions $f(x)$ and $f_{LL}(x)$ are symmetric around $x=1/2$, it therefore follows from Eqs.~\eqref{sumft} that we obtain the following results for the momentum sum rule of the valence quark: 
\begin{align}
\int_0^1 \, x \, dx \, f(x) &= \frac{1}{2}, &
\int_0^1 \, x \, dx \, f_{LL}(x) &= 0.  \label{msumft}
\end{align}
Because the valence antiquark gives the same contribution to the momentum sum, this means that together the valence quark and antiquark carry 100$\%$ of the light-cone momentum, independently of the spin quantization axis $\vect{S}$ and spin projection $\l$ of the hadron.

The sum rules for the PDFs $g(x)$ and $h(x)$ [see Eqs.~\eqref{pdf2} and \eqref{pdf3}] are  conventionally called the quark spin sum and the tensor charge, respectively, and are model dependent. More precisely, our quantity $\int_0^1 dx \, g(x)$ is  {\em twice} the expectation value of the spin operator of the valence {\em quark} in a longitudinally polarized $\rho$-meson. In our model the valence antiquark gives the same contribution, it therefore follows that $\int_0^1 dx \, g(x)$ is the sum of the valence quark and antiquark spins. The ``naive'' (nonrelativistic)  quark model result for this sum rule would equal the physical spin of the $\rho$ meson, namely unity.\footnote{For the case of a spin-$1/2$ hadron (e.g. a nucleon), it is common to define the ``naive'' quark model result for the sum rule to be unity, which is {\em twice} the physical spin of the nucleon.} However, relativistic effects can lead to substantial reductions  of this sum rule because of non-zero quark orbital angular momentum. Similar comments also apply for the tensor charge, which in the nonrelativistic quark model would take the same value as the spin sum (unity in this case), and for the tensor charge relativistic effects tend to give smaller reductions than to the spin sum~\cite{Jaffe:1991ra}. 

%===============================================================================
%===============================================================================
\section{RESULTS\label{results}}
This section presents results for the $\rho^+$ meson TMDs and PDFs. In the calculation there appear three 
model parameters, namely the dressed quark mass $M$, and two proper-time regularization parameters $\L_{IR}$ and $\L_{UV}$. 
Following previous work~\cite{Cloet:2007em,Cloet:2014rja,Carrillo-Serrano:2016igi,Carrillo-Serrano:2015uca} we set 
$M=0.4\,$GeV and  $\L_{IR} = 0.24\,$GeV, where the motivation for the latter is that it should be of order $\L_{\rm QCD}$ 
because it introduces the confinement scale into our calculations. The ultraviolet cutoff $\L_{UV}$ is then fit to 
reproduce the experimental pion decay constant $f_{\pi} = 0.093$ GeV, which gives $\L_{UV} = 0.645\,$GeV. For the 
$\rho$-meson mass we take the physical value of $m_\rho = 0.776\,$GeV, and using Eq.~\eqref{grho} this gives a 
$\rho$ meson--quark-antiquark coupling constant of $Z_\rho = 2.61$. We note that the physical $\rho$ mass is only 
slightly smaller than $2\,M=0.8\,$GeV, however because our infrared cut-off $\L_{IR}$ eliminates this unphysical 
threshold for $\rho$ decay into dressed quarks, the qualitative behavior of all our results remains unchanged even 
if lower values of $M$ are chosen.

%=====================================================================
\begin{figure}[h!]
\centering
\includegraphics[width=\columnwidth]{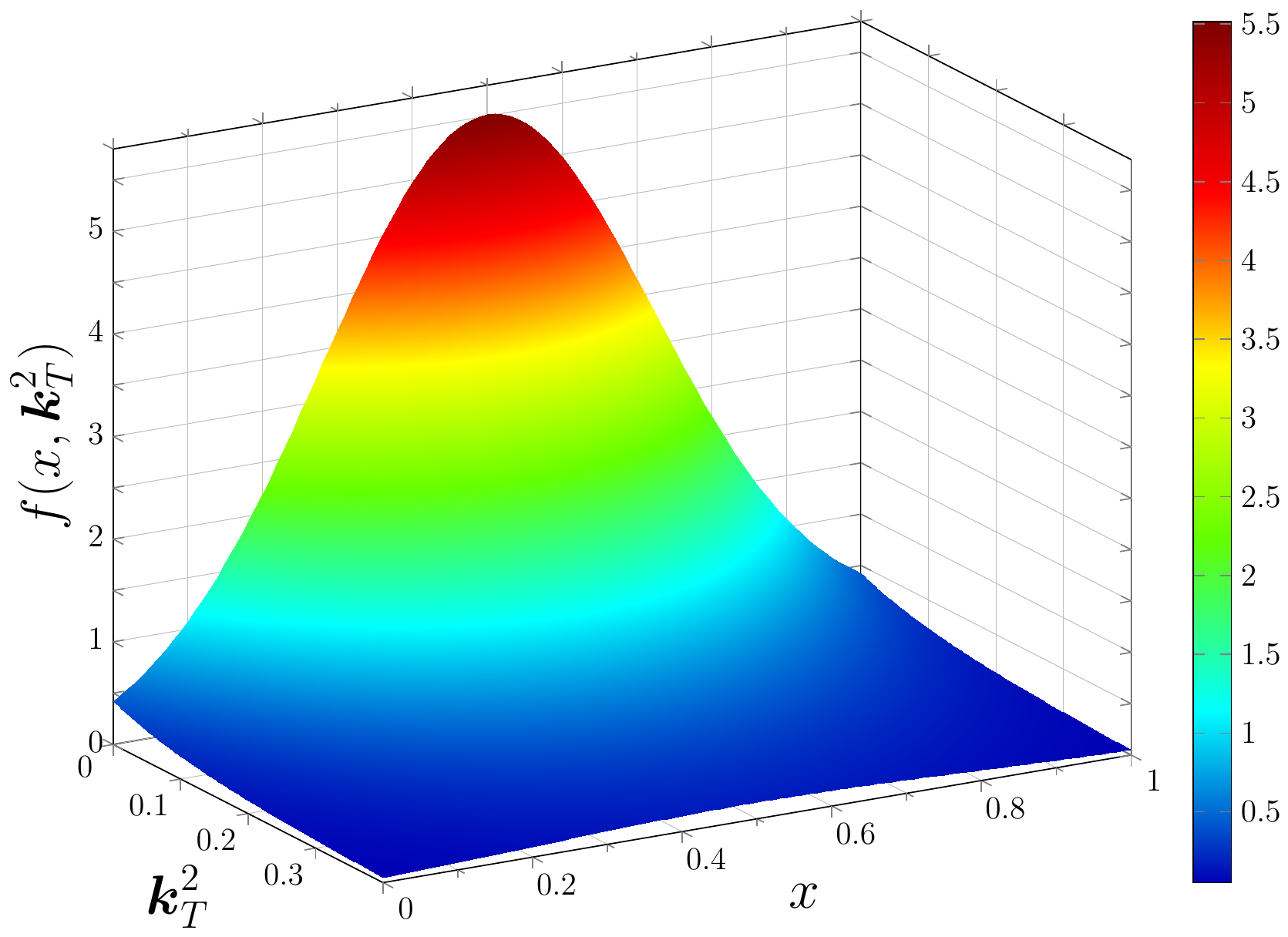}  \\
\includegraphics[width=\columnwidth]{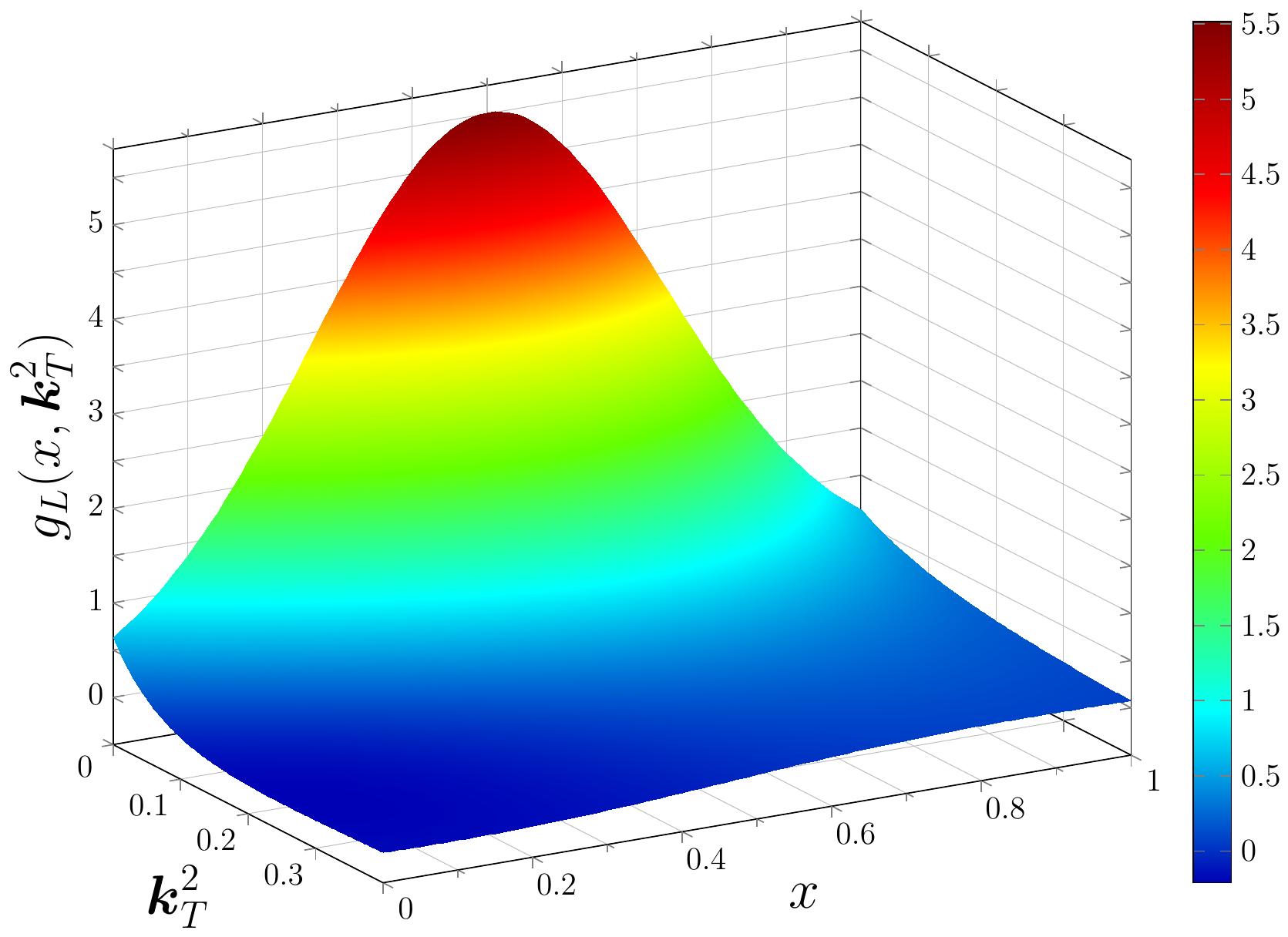}  \\
\includegraphics[width=\columnwidth]{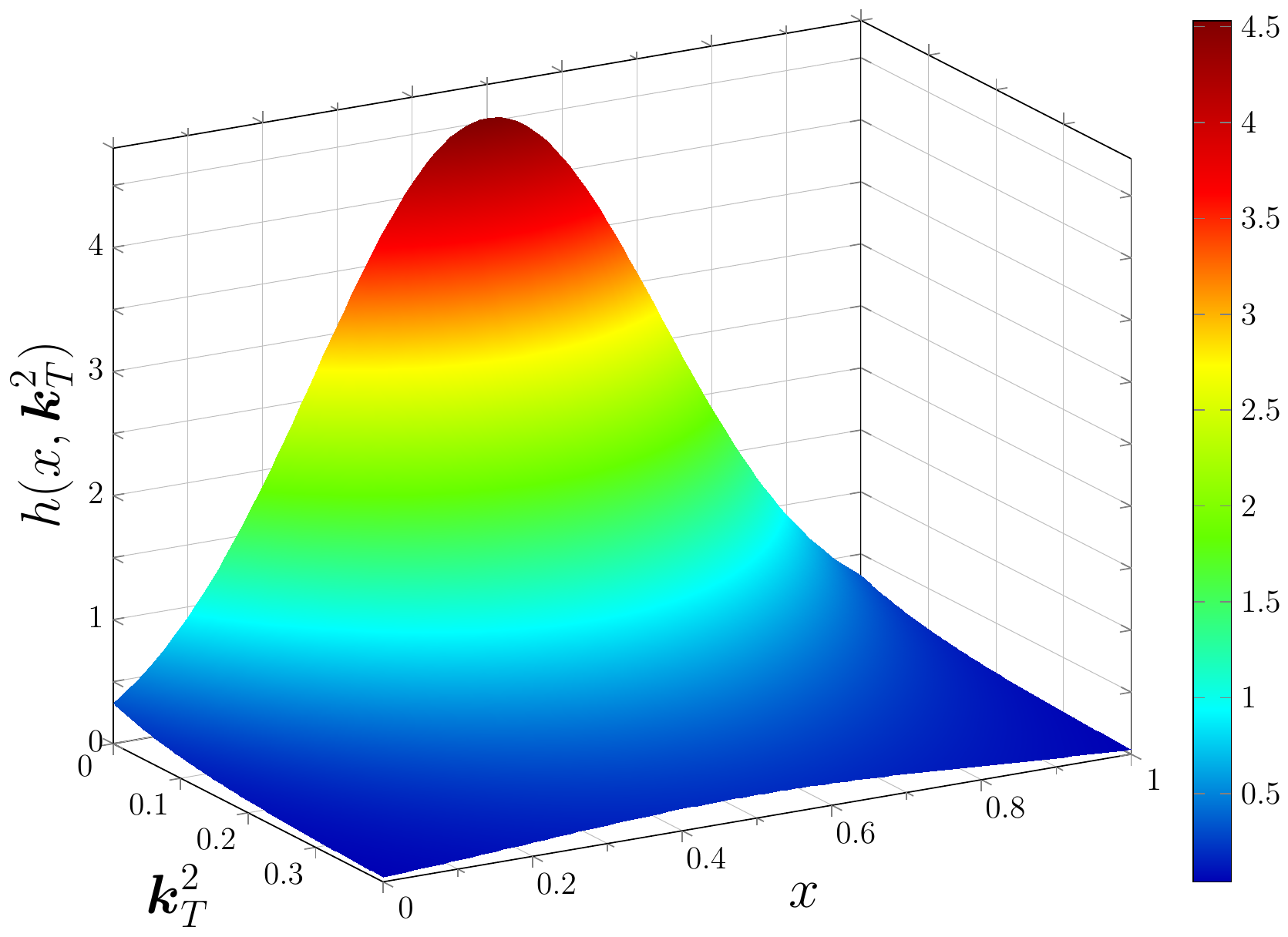} 
\caption{(Color online) Results for the TMDs $f(x,\vect{k}_T^2)$, $g_{L}(x,\vect{k}_T^2)$ and $h(x,\vect{k}_T^2)$ for the $\rho^+$ meson. These TMDs are defined in Eqs.~\eqref{form1}, \eqref{form2} and  \eqref{form3}, and when integrated over $\vect{k}_T$ give the PDFs 
$f(x)$, $g(x)$ and $h(x)$ which are 
familiar from the spin-half case. 
The TMDs are given in units of GeV$^{-2}$ and $\vect{k}_T^2$ in GeV$^{2}$.
}
\label{fig:tmd1}
\end{figure}
%=====================================================================

%===============================================================================
%===============================================================================
\subsection{Rho meson TMDs}
Among the nine TMDs which appear in Eqs.~\eqref{form1}--\eqref{form3} the pretzelosity 
TMD, $h_T^{\perp}(x, \vect{k}_T^2)$, vanishes identically in our calculations. This leaves eight non-vanishing 
TMDs. Results for the three TMDs $f(x,\vect{k}_T^2)$, $g_L(x,\vect{k}_T^2)$ and 
$h(x,\vect{k}_T^2)$ -- which 
when integrated over $\vect{k}_T$ give the PDFs $f(x)$, $g(x)$ and $h(x)$ familiar from the spin-half case -- 
are given in Fig.~\ref{fig:tmd1}. We find that all three TMDs are of similar magnitude at $\vect{k}_T^2 \simeq 0$, 
with each reaching a maximum at $x = 1/2$. The TMDs $f(x, \vect{k}_T^2)$ and $h(x, \vect{k}_T^2)$ are symmetric 
under the transformation $x \to 1-x$, while $g_L(x, \vect{k}_T^2)$ has this symmetry only at $\vect{k}_T^2 = 0$. 
As the quark transverse momentum $\vect{k}_T^2$ grows all three TMDs develop a much weaker dependence on the 
light-cone momentum fraction $x$, which indicates that a struck quark with large $\vect{k}_T^2$ has a much 
weaker preference for any particular $x$ value. For the helicity TMD we find that as $\vect{k}_T^2$ gets 
larger an increasing fraction of $g_L(x,\vect{k}_T^2)$ becomes negative, beginning in the low $x$ region 
and moving to higher $x$ as $\vect{k}_T^2$ increases. This indicates that at small $\vect{k}_T^2$ the quark 
and hadron spins dominantly align, but as $\vect{k}_T^2$ gets larger anti-alignment of quark and hadron spins 
becomes increasingly important, starting from the low $x$ region. 
Explicit expressions for these TMDs are given in Eqs.~\eqref{eq:f1}, \eqref{eq:gL1} and \eqref{eq:h1}, and when viewed as functions of $\vect{k}_T$ our results do not take a Gaussian form. This is discussed further below.
%
%Viewed as functions of $\vect{k}_T$, all 
%three TMDs are essentially Gaussian functions, which is a consequence of the proper-time regularization scheme, 
%as can be seen in the explicit expressions given in Appendix~\ref{ex-form}.    

%=====================================================================
\begin{figure}[tbp]
\centering\includegraphics[width=\columnwidth]{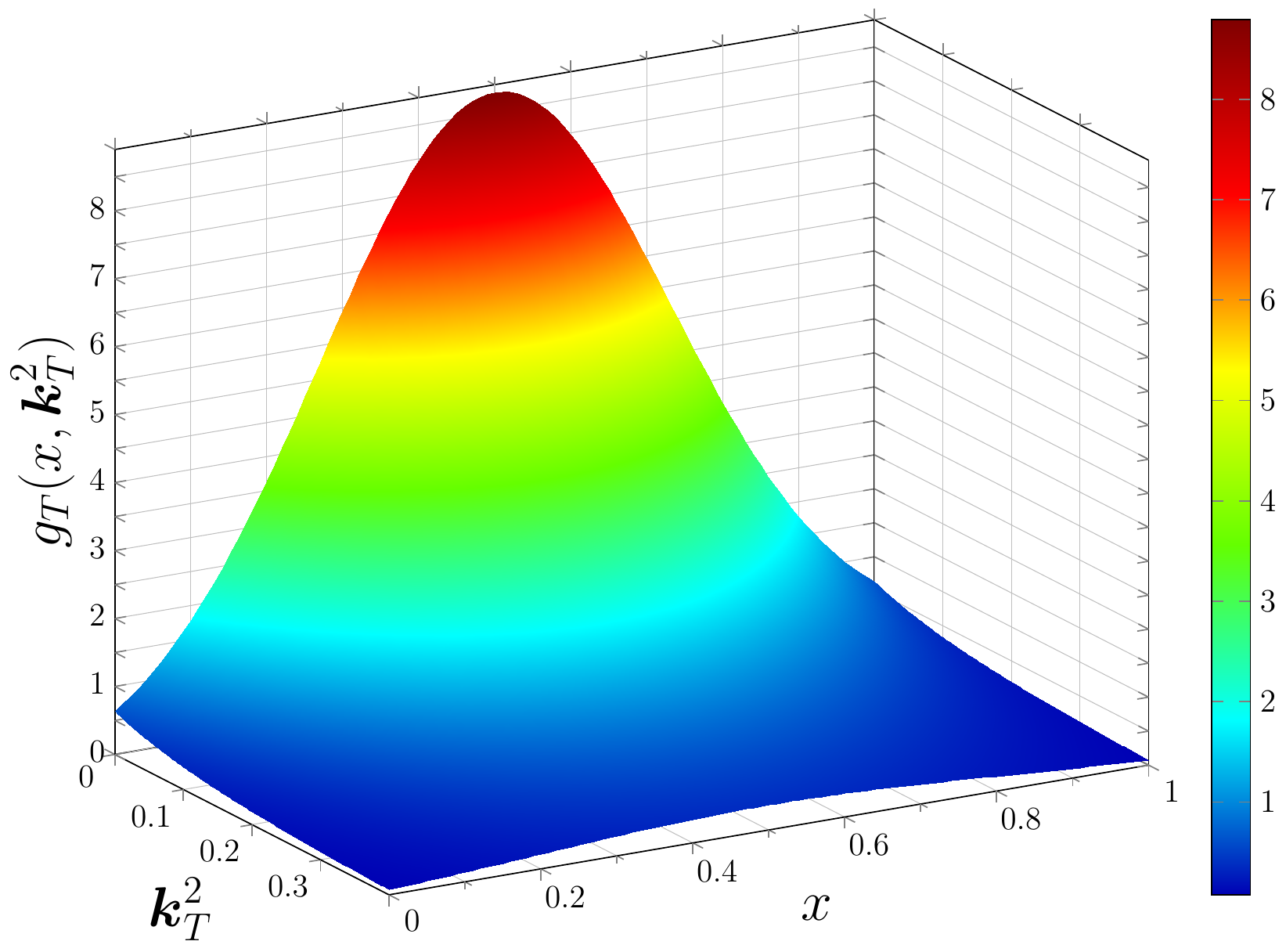} \\
\centering\includegraphics[width=\columnwidth]{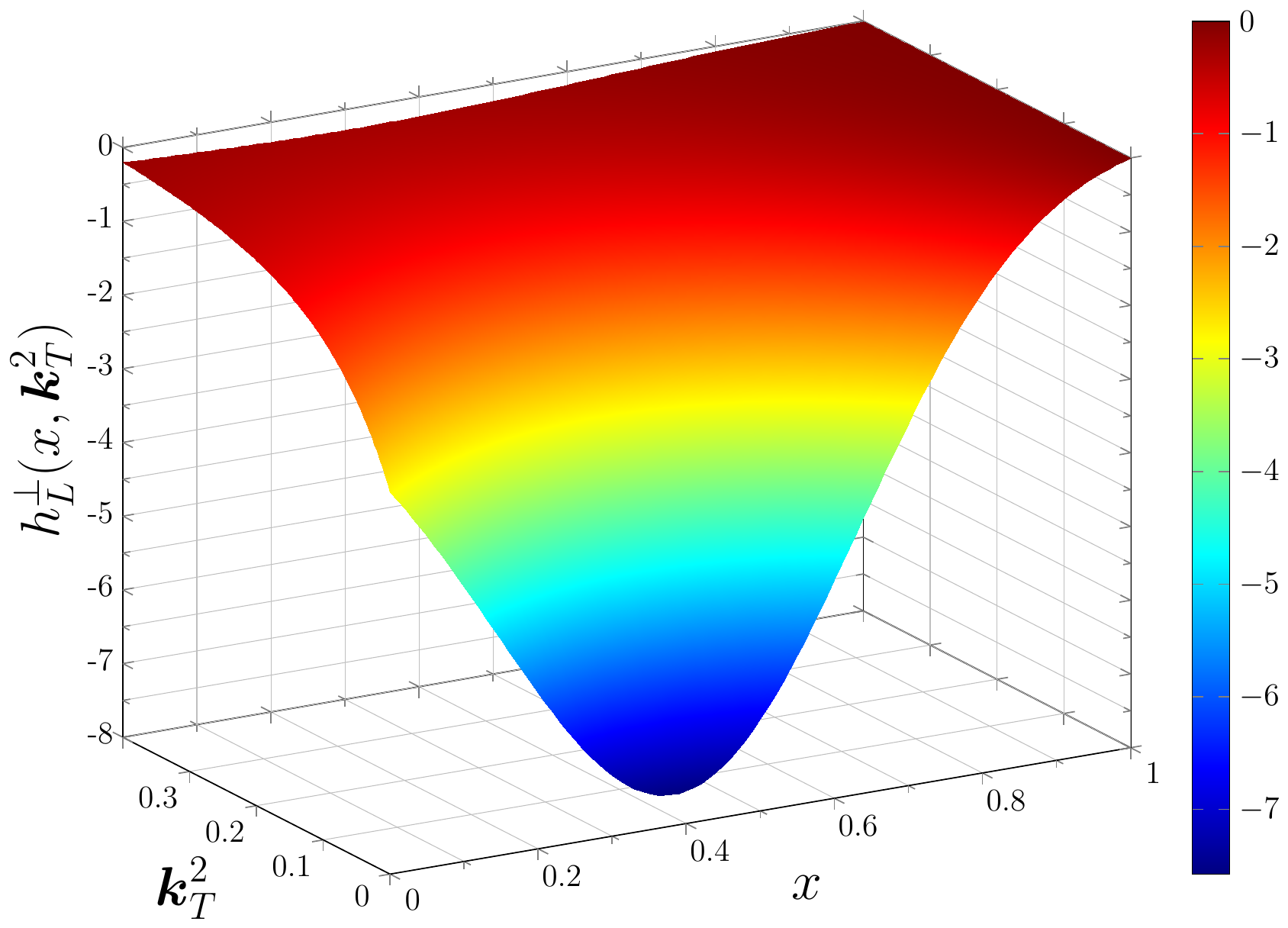} 
\caption{(Color online) Results for $g_T(x,\vect{k}_T^2)$ and $h_{L}^{\perp} (x,\vect{k}_T^2)$ of a $\rho^+$ meson. These TMDs are defined in Eqs.~\eqref{form2} and \eqref{form3}, where the pretzelosity TMD $h_T^{\perp}(x, \vect{k}_T^2)$ vanishes identically in our calculations. The TMDs are given in units of GeV$^{-2}$ and $\vect{k}_T^2$ in GeV$^{2}$.}
\label{fig:tmd2}
\end{figure}
%=====================================================================

Results for the TMDs $g_T(x,\vect{k}_T^2)$ and $h_{L}^{\perp}(x,\vect{k}_T^2)$ are illustrated in Fig.~\ref{fig:tmd2}, with explicit expressions given in Eqs.~\eqref{eq:gT1} and \eqref{eq:hL}. We find that at $\vect{k}_T^2\simeq 0$ each of these TMDs has a magnitude about twice that of $f(x, \vect{k}_T^2)$. However, one should keep in mind that the TMDs shown in Fig.~\ref{fig:tmd2} are 
multiplied by a factor $\simeq \vect{k}_T/m_{\rho}$ when they appear in observables, which is evident from 
Eqs.~\eqref{form2} and \eqref{form3}. The TMD $g_T(x,\vect{k}_T^2)$ is a positive function with a peak at 
$x=1/2$ and is symmetric under the transformation $x \to 1-x$, and we find it is related to the transversity 
TMD by
\begin{align}
g_T(x,\vect{k}_T^2) = \frac{m_\rho}{M}\,h(x,\vect{k}_T^2).
\end{align}
However, this relation is likely only valid at the model scale and would not survive QCD evolution. 
The TMD $h_{L}^{\perp}(x,\vect{k}_T^2)$ is a negative function with no definite symmetry in $x$, and 
a peak in magnitude at $x\simeq 0.4$ when $\vect{k}_T^2 \simeq 0$ which moves to smaller values of $x$ as 
$\vect{k}_T^2$ gets larger. As $\vect{k}_T^2$ becomes larger we again find that $g_T(x,\vect{k}_T^2)$ and 
$h_{L}^{\perp}(x,\vect{k}_T^2)$ develop a much weaker dependence on the light-cone momentum fraction $x$, 
and as a function of $\vect{k}_T$ these TMDs do not take a Gaussian form.
%
%essentially behave as Gaussian functions (see Appendix~\ref{ex-form}). 
%
Interestingly, our result for $h_{L}^{\perp}(x,\vect{k}_T^2)$, as well as $h(x,\vect{k}_T^2)$ illustrated 
in Fig.~\ref{fig:tmd1}, are both proportional to the dressed quark mass $M$ (see Appendix~\ref{ex-form}). 
These TMDs may therefore be considered to arise from the mechanism of dynamical chiral symmetry breaking. 

%=====================================================================
% \begin{figure*}[tbp]
% %
% \centering\includegraphics[width=\columnwidth]{plot_tmd_rho_thetaLL_alt2}  \hfill
% \centering\includegraphics[width=\columnwidth]{plot_tmd_rho_thetaTT_alt1}  \\[1.0ex]
% \centering\includegraphics[width=\columnwidth]{plot_tmd_rho_thetaLT_alt2}  \hfill
% \centering\includegraphics[width=\columnwidth]{plot_tmd_rho_theta_alt2}
% %
% \caption{(Color online) Results for the tensor polarized TMDs $f_{LL} (x,\vect{k}_T^2)$, 
% $f_{LT} (x, \vect{k}_T^2)$, and
% $f_{TT} (x,\vect{k}_T^2)$ for the $\rho^+$ meson. These TMDs are defined in Eqs.~\eqref{form1}, 
% and are given in units of GeV$^{-2}$ and $\vect{k}_T^2$ in GeV$^{2}$.}
% \label{fig:tmd3}
% \end{figure*}
%=====================================================================

%=====================================================================
\begin{figure}[tbp]
\centering\includegraphics[width=\columnwidth]{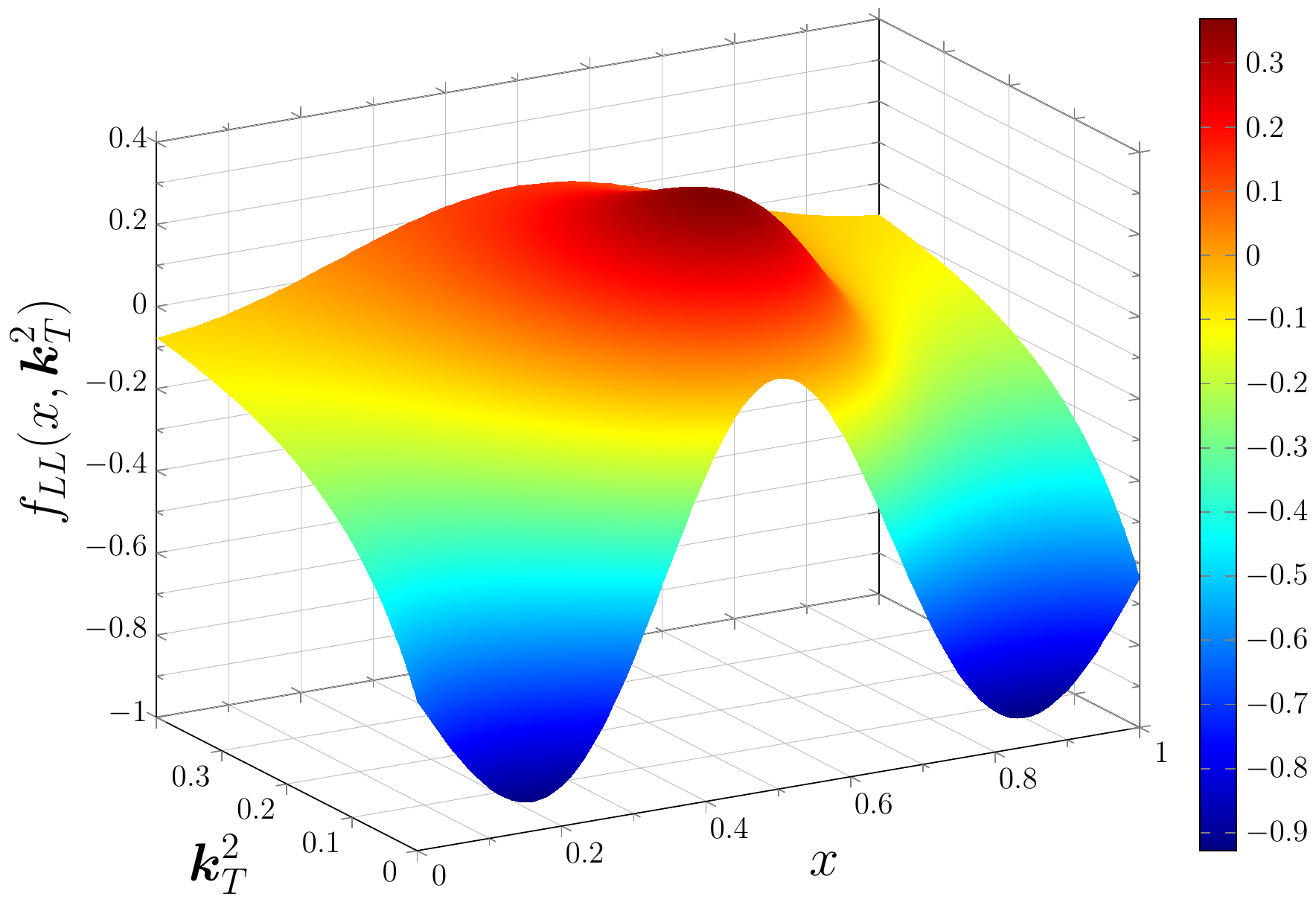}  \\
\centering\includegraphics[width=\columnwidth]{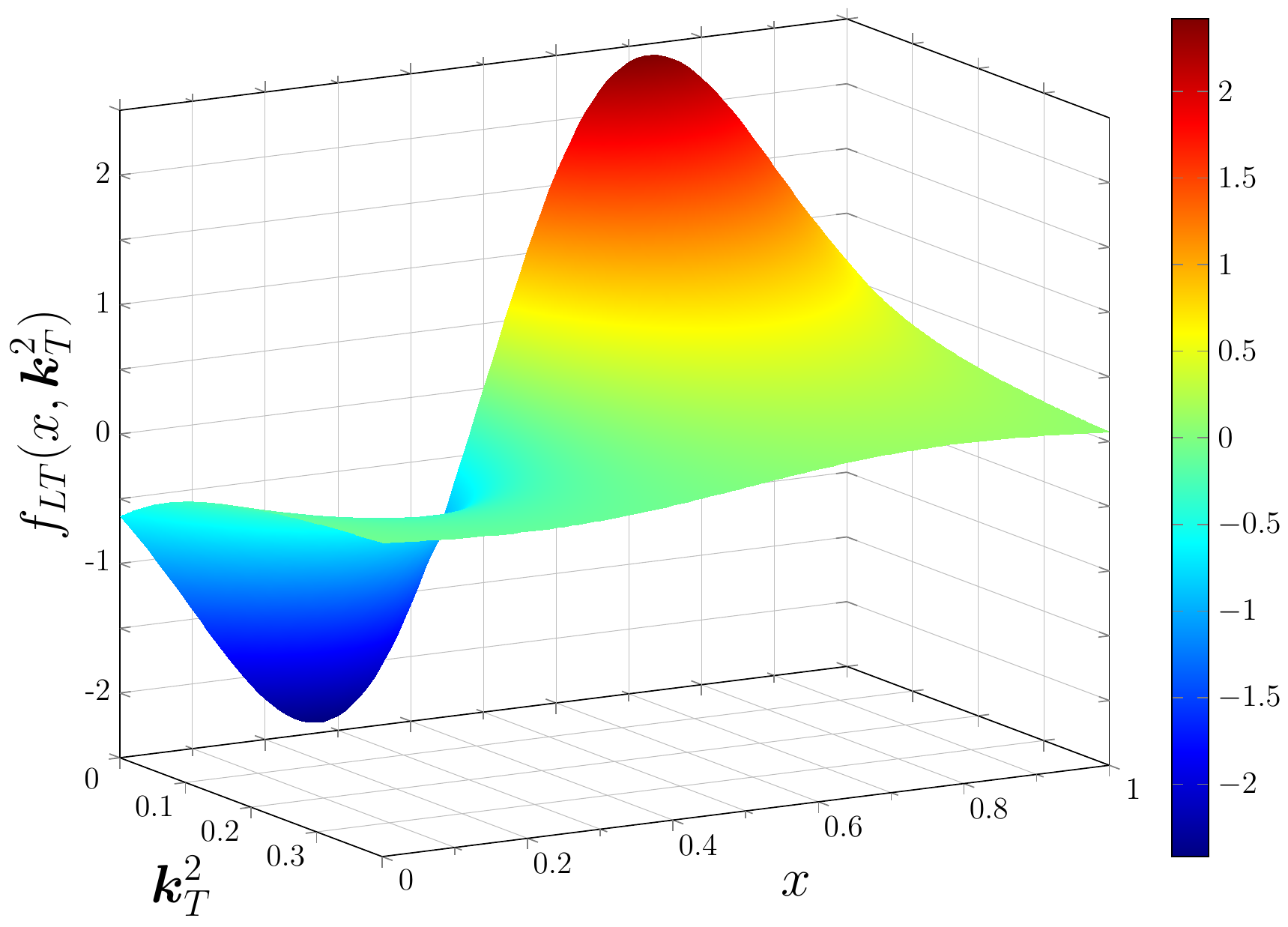} \\
\centering\includegraphics[width=\columnwidth]{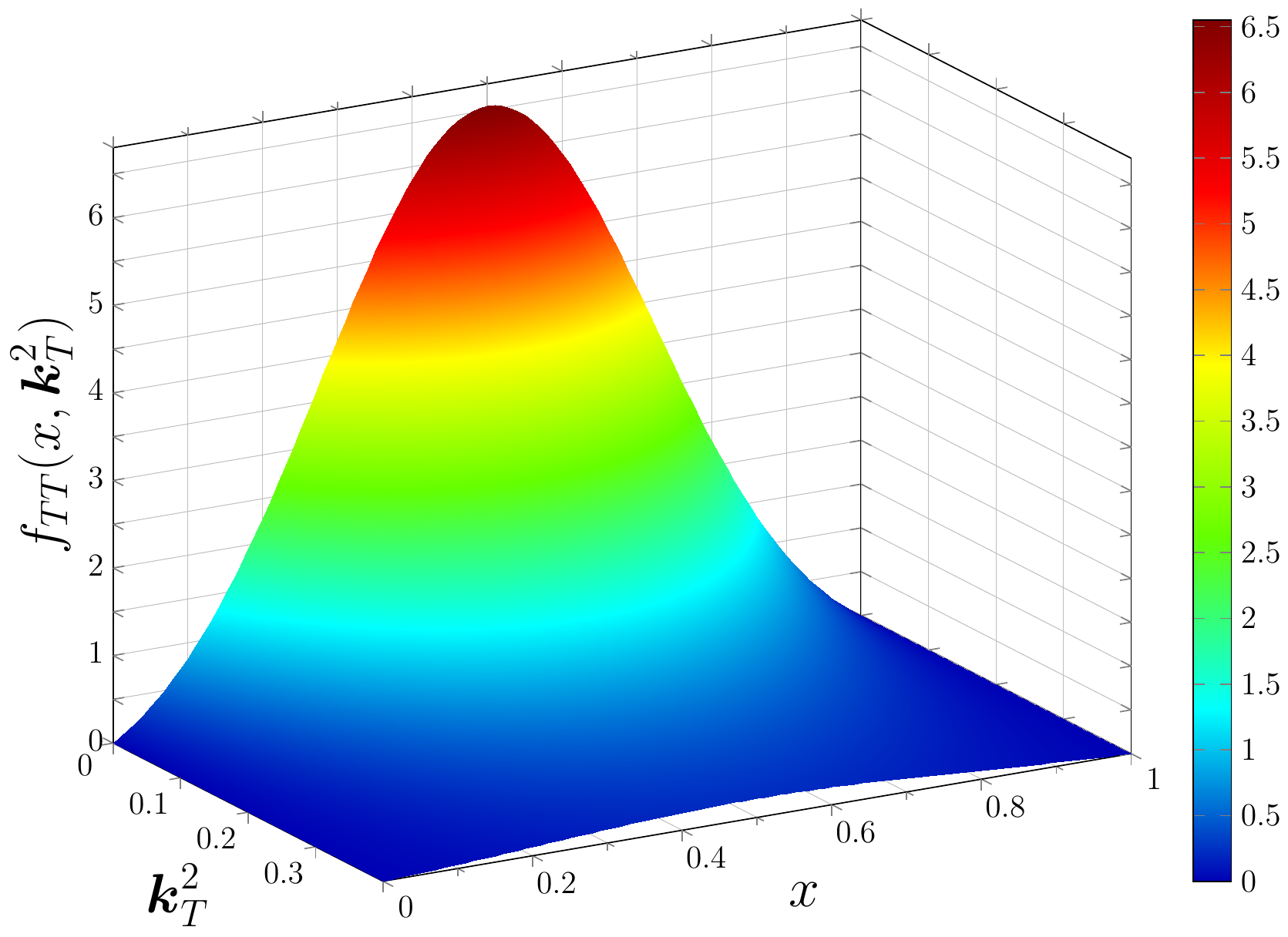}
\caption{(Color online) Results for the tensor polarized TMDs $f_{LL} (x,\vect{k}_T^2)$, 
$f_{LT} (x, \vect{k}_T^2)$, and
$f_{TT} (x,\vect{k}_T^2)$ for the $\rho^+$ meson. These TMDs are defined in Eqs.~\eqref{form1}, 
and are given in units of GeV$^{-2}$, with $\vect{k}_T^2$ in GeV$^{2}$.}
\label{fig:tmd3}
\end{figure}
%=====================================================================

%=====================================================================
\begin{figure}
\centering\includegraphics[width=\columnwidth]{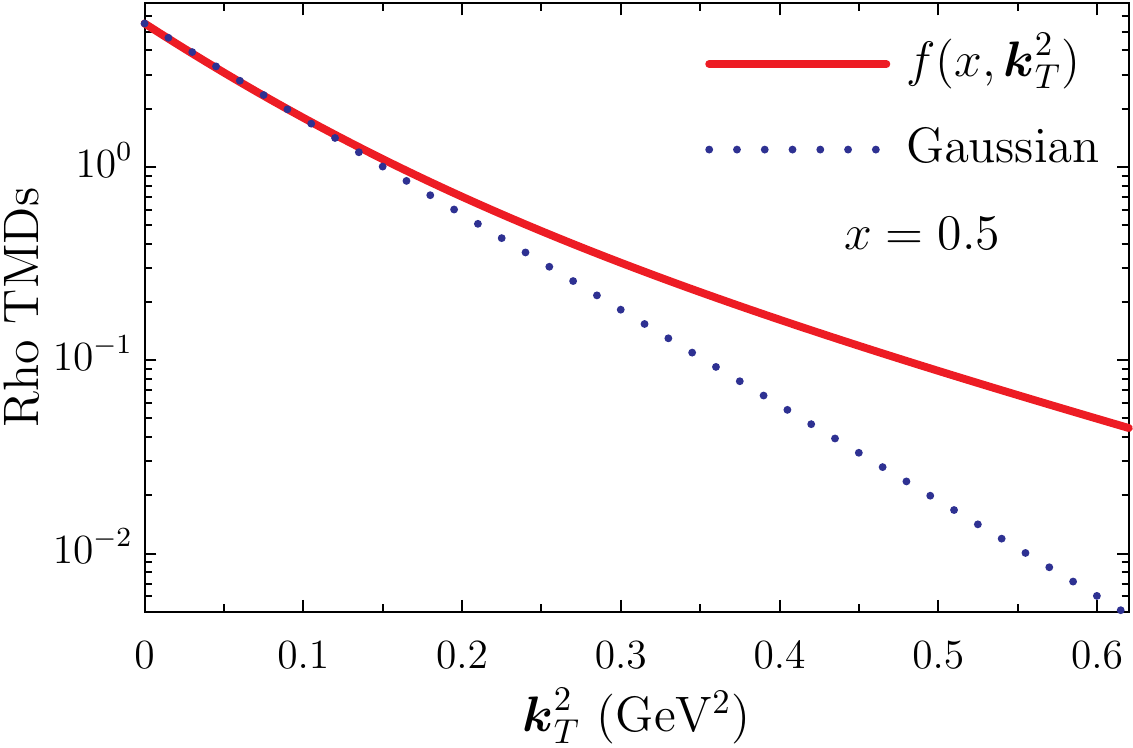}
\caption{(Color online) The solid line is the TMD $f(x,\vect{k}_T^2)$ for $x=0.5$ 
and the dotted line is a Gaussian fit, of the form given in Eq.~\eqref{eq:gaussian}, to this result.}
\label{fig:xslices}
\end{figure}
%=====================================================================

Results for the three $\rho^+$ tensor polarized TMDs $f_{LL} (x,\vect{k}_T^2)$,  $f_{LT} (x,\vect{k}_T^2)$, and $f_{TT} (x,\vect{k}_T^2)$, which do not appear for spin-half hadrons,  are illustrated in Fig.~\ref{fig:tmd3}. We find that $f_{TT} (x,\vect{k}_T^2)$ has similar characteristics  to the other TMDs that also appear for the spin-half case, that is, a peak at $x=1/2$ and a  symmetry under the transformation $x \to 1-x$.  The TMD $f_{LL}(x,\vect{k}_T^2)$ is symmetric under the  transformation $x \to 1 - x$, whereas $f_{LT} (x,\vect{k}_T^2)$ is antisymmetric. 
Interestingly, $f_{LT}(x, \vect{k}_T^2)$ vanishes identically at $x=1/2$ for all $\vect{k}_T^2$. This vanishing along the 
line $x=1/2$ is a signal that this TMD is very sensitive to components in the $\rho^+$ wave function that 
have non-zero quark orbital angular momentum. That is, $s$-wave components tend to peak at zero relative 
momentum, which corresponds to $x=1/2$, and since 
$f_{LT}(x,\vect{k}_T^2)$ vanishes along this 
line but is in general non-zero, it must receive significant contributions from components of the 
$\rho^+$ wave function with $L \geqslant 1$. 
%
%These findings confirm earlier analysis which found that the 
%tensor polarized structure function, $b_1(x)$, is a sensitive measure of quark orbital angular 
%momentum~\cite{Hoodbhoy:1988am,Close:1990zw,Umnikov:1996qv}.
%The discussion in Sect.~\ref{form-Nino} (see also Appendix~\ref{app:relations}) 
%demonstrates that the TMD $\theta_{LL} (x,\vect{k}_T^2)$ does not independently appear in %observables such as cross-sections 
%and asymmetries, but only in combination with $\theta_{TT}(x,\vect{k}_T^2)$. This combination is %given by 
%Eq.~\eqref{theta_tmd}, which defines the TMD $\theta (x,\vect{k}_T^2)$, that when integrated over 
%$\vect{k}_T^2$ is proportional to $b_1(x)$ [see Eqs.~\eqref{theta} and \eqref{bt}]. 
%
The TMD $f_{LL}(x,\vect{k}_T^2)$, however, instead of vanishing along the line $x=1/2$ as is the case for $f_{LT}(x, \vect{k}_T^2)$, vanishes along two curves in the  $(x,\,\vect{k}_T^2)$-plane which are symmetric about the line $x=1/2$, but have a strong dependence on the quark transverse momentum. This behavior, too, is a signal of non-zero quark orbital angular momentum. Viewed as a function of $\vect{k}_T$, the TMDs $f_{LL}(x,\vect{k}_T^2)$, $f_{LT}(x,\vect{k}_T^2)$ and $f_{TT}(x,\vect{k}_T^2)$ do not behave as Gaussian functions.
%
%however $f_{LL}(x,\vect{k}_T^2)$ has a behavior not seen in the  other tensor polarized TMDs. 
%
For example, in the neighborhood of the $x=1/2$ line $f_{LL}(x,\vect{k}_T^2)$ vanishes at $\vect{k}_T^2 = 0$, then increases as $\vect{k}_T^2$ grows reaching  a peak near $\vect{k}_T^2 = 0.1\,$GeV$^2$, and then decreases slowly as $\vect{k}_T^2$ gets larger.  This interesting behavior of $f_{LL}(x,\vect{k}_T^2)$, which is intimately linked to quark orbital angular  momentum, makes it an important TMD for further study and measurement. In this connection we note that while $f_{LT} (x,\vect{k}_T^2)$ is up to two times larger than  $f_{LL}(x,\vect{k}_T^2)$, it enters with a factor $\simeq \vect{k}_T/m_{\rho}$, making $f_{LL}(x,\vect{k}_T^2)$ dominant in the small transverse momentum region.

In phenomenological applications~\cite{Anselmino:2013lza,Echevarria:2014xaa,DAlesio:2014mrz,Bacchetta:2017gcc,Scimemi:2017etj} it common to use a separable Gaussian representation for a general TMD which takes the form
\begin{align}
q(x,\vect{k}_T^2) = q(x)\ \frac{e^{-\vect{k}_T^2/\la k_T^2\ra}}{\pi\,\la k_T^2\ra},
\label{eq:gaussian}
\end{align}
where $\la k_T^2\ra$ is the average $\vect{k}_T^2$. It is clear from our explicit results for the TMDs, given in Appendix~\ref{ex-form}, that they are neither separable or of a Gaussian form. Performing the integration over proper-time we find TMDs with a $\vect{k}_T^2$-dependence which is a sum of terms of the form
\begin{align}
%q(x,\vect{k}_T^2) \propto \sum_{n=1,2}\,q_n(x)\, \frac{e^{-\lf[\vect{k}_T^2 + B\rg]/\L_{UV}^2} - e^{-\lf[\vect{k}_T^2 + B\rg]/\L_{IR}^2}}{\lf[\vect{k}_T^2 + B\rg]^{n}},
q(x,\vect{k}_T^2) \propto \lf[\vect{k}_T^2\rg]^\a \,\frac{e^{-\lf[\vect{k}_T^2 + B\rg]/\L_{UV}^2} - e^{-\lf[\vect{k}_T^2 + B\rg]/\L_{IR}^2}}{\lf[\vect{k}_T^2 + B\rg]^{n}},
\end{align}
where $B=M^2 - x(1-x)\,m_\rho^2$, $n=1,2$ and $\a=0,1$. Therefore we find a difference of Gaussian's with an additional power-law behavior in $\vect{k}_T^2$. As an illustrative example in Fig.~\ref{fig:xslices} we present $f(x,\vect{k}_T^2)$ for $x=0.5$ and contrast this with a Gaussian fit. We find that the Gaussian only provides a reasonable description for $\vect{k}_T^2 \leqslant 0.2$, with the complete result falling off slower than the Gaussian over the range illustrated. In fact, a Gaussian form can only be obtained for some TMDs where $\a=0$, and in the limits $\vect{k}_T^2 \ll B$ and $\L_{IR} \to 0$, where the latter limit removes quark confinement in our calculations.

%=====================================================================
\begin{figure*}[tbp]
\centering\includegraphics[width=1.00\columnwidth]{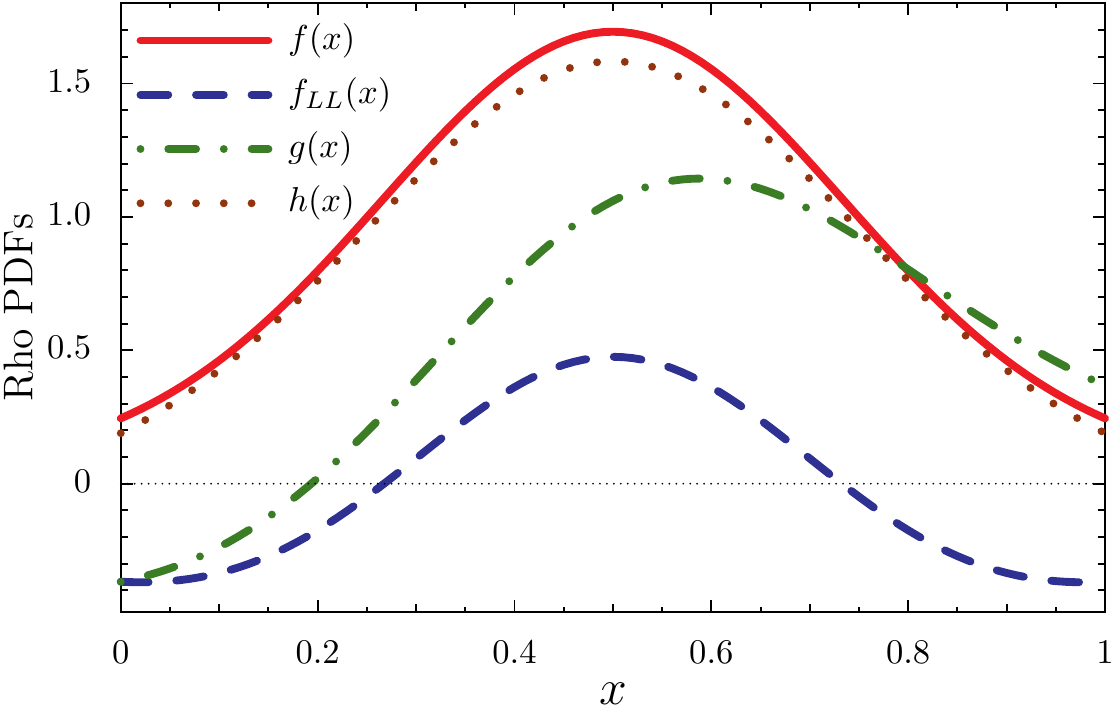} \hfill
\centering\includegraphics[width=1.03\columnwidth]{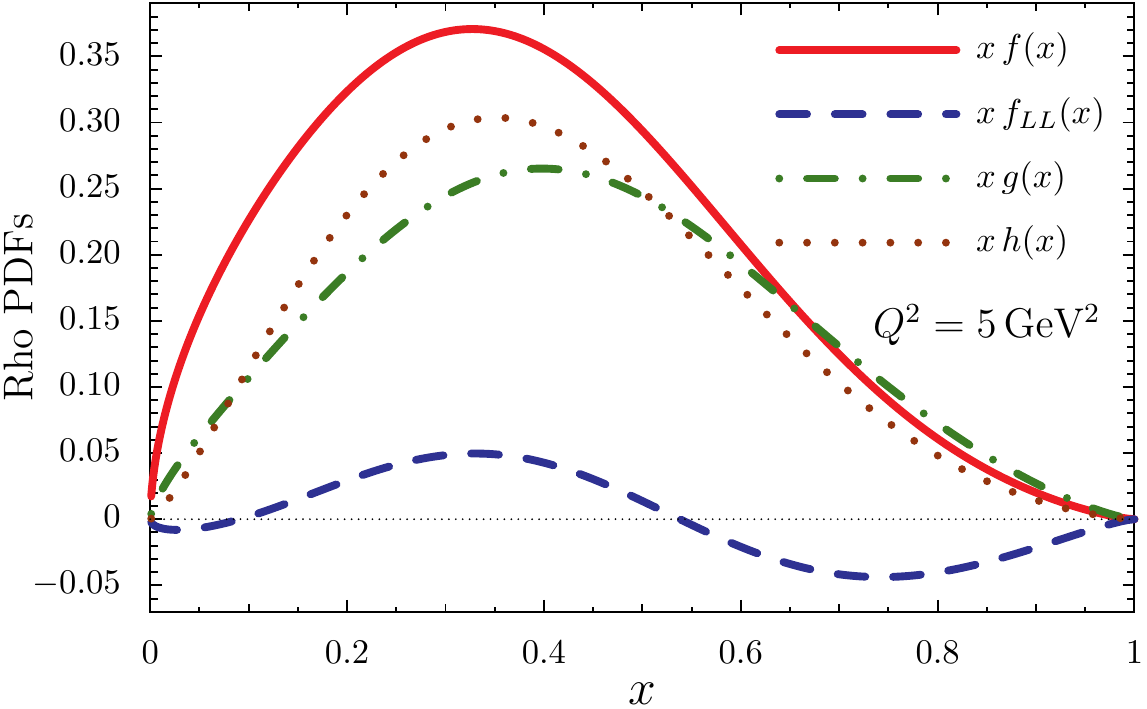}
\caption{(Color online) We illustrate results for the unpolarized, helicity, transversity and tensor polarized PDFs of the 
$\rho^+$ meson. The left panel gives results at the model scale $Q_0^2 = 0.16$ GeV$^2$ and the right panel 
presents the PDF results, multiplied by $x$, at the scale $Q^2 = 5\,$GeV$^2$. The NLO non-singlet DGLAP evolution was performed using the codes of Refs.~\cite{Miyama:1995bd,Hirai:1997gb,Hirai:1997mm}.}
\label{fig:pdfs}
\end{figure*}
%=====================================================================

%===============================================================================
% \begin{table}[tbp]
% \addtolength{\tabcolsep}{6.7pt}
% \addtolength{\extrarowheight}{2.2pt}
% \begin{tabular}{ccccccccc}
% \hline\hline
%                & $f$  & \ph{-}$g_L$  & $g_T$ & $h$ & $h_L^\perp$ & $f_{TT}$ \\ [0.2em]
% \hline
% $\lf<k_T\rg>$  & 0.41 & \ph{-}0.10   & 0.44 & 0.44 & 0.42  & 0.42           \\
% %
% $\la k_T^2\ra$ & 0.22 & -0.18        & 0.27 & 0.27 & 0.24  & 0.24           \\[0.3em]
% \hline\hline
% \end{tabular}
% \caption{The $\la k_T\ra$ and $\la k_T^2\ra$ moments of the various TMDs as defined by Eq.~\eqref{eq:kT}.
% Note, the denominator of Eq.~\eqref{eq:kT} vanishes for the TMDs $f_{LL}(x,\vect{k}_T^2)$ and $f_{LT}(x,\vect{k}_T^2)$, so they are not listed.}
% \label{tab:kT}
% \end{table}
%===============================================================================

%===============================================================================
\begin{table}[tbp]
\addtolength{\tabcolsep}{6.7pt}
\addtolength{\extrarowheight}{2.2pt}
\begin{tabular}{ccccccccc}
\hline\hline
               & $f$  & \ph{-}$g_L$  & $g_T$ & $h$ & $h_L^\perp$ & $f_{TT}$ \\ [0.2em]
\hline
$\lf<k_T\rg>$  & 0.32 & \ph{-}0.08   & 0.34 & 0.34 & 0.33  & 0.32           \\
$\la k_T^2\ra$ & 0.13 & -0.11        & 0.16 & 0.16 & 0.15  & 0.14           \\[0.3em]
\hline\hline
\end{tabular}
\caption{The $\la k_T\ra$ and $\la k_T^2\ra$ moments of the various TMDs as defined by Eq.~\eqref{eq:kT}, in units of GeV and GeV$^2$ respectively. Note, the denominator of Eq.~\eqref{eq:kT} vanishes for the TMDs $f_{LL}(x,\vect{k}_T^2)$ and $f_{LT}(x,\vect{k}_T^2)$, so they are not listed.}
\label{tab:kT}
\end{table}
%===============================================================================

The transverse-momentum dependence of the TMDs can be further revealed through $\vect{k}_T$-weighted moments, such as
\begin{align}
\la k_T^n \ra_\a \equiv \frac{\int_0^1 dx \int d^2\vect{k}_T\,\lf|\vect{k}_T\rg|^n \ \a(x,\vect{k}_T^2)}{\int_0^1 dx \int d^2\vect{k}_T \, \a(x,\vect{k}_T^2)},
\label{eq:kT}
\end{align}
for an arbitrary TMD $\a(x,\vect{k}_T^2)$. Results for $\lf<k_T\rg>$ and $\la k_T^2\ra$ for the various TMDs are given in Tab.~\ref{tab:kT}. Typical values are $\la k_T^2\ra \sim 0.15\,$GeV$^2$, which is of a similar magnitude to the square of the dressed quark mass $M$.
%however for $g_L(x,\vect{k}_T^2)$ we find a much smaller value, which reflects its sign changes in the $(x,\vect{k}_T^2)$-plane.
Tab.~\ref{tab:kT} also makes clear that the $\vect{k}_T$-weighted moments of the TMDs with no connection to PDFs need not vanish, even though they do not contribute to the quark correlator of Eq.~\eqref{phi1} when it is integrated over $\vect{k}_T$.

%===============================================================================
%===============================================================================
\subsection{Rho meson PDFs}
In Fig.~\ref{fig:pdfs} we illustrate results for the PDFs $f(x)$, $g(x)$, and $h(x)$, which are familiar 
from the spin-half case, as well as the tensor polarized PDF $f_{LL}(x)$ specific to hadrons with spin 
$J \geqslant 1$. The left panel shows results at the model scale, which we set as $Q_{0}^2 = 0.16$ GeV$^{2}$ 
following previous studies~\cite{Cloet:2007em}. The right panel presents results obtained by performing the NLO 
non-singlet QCD evolution to 
$Q^2 = 5\,$GeV$^2$, where we used the DGLAP evolution codes of Refs.~\cite{Miyama:1995bd,Hirai:1997gb,Hirai:1997mm}. 

At the model scale, we find that the unpolarized and transversity PDFs have very similar behaviors, 
whereas the helicity PDF is suppressed relative to $f(x)$ below about $x \sim 0.75$, but becomes larger than $f(x)$ at higher $x$. For spin-half 
hadrons $g(x)$ must always be smaller in magnitude than $f(x)$ because of the positivity constraint $|g(x)| \leqslant f(x)$. 
However, for spin-one hadrons this positivity constraint is modified [see Eq.~\eqref{bound_PDF-3}] and 
therefore our result in  Fig.~\ref{fig:pdfs}, where  $|g(x)| > f(x)$ at large $x$, does not necessarily 
violate positivity. In fact, as we shall see, our results satisfy Eq.~\eqref{bound_PDF-3}.
After QCD evolution to $Q^2 = 5$ GeV$^{2}$, the transversity PDF becomes suppressed relative to $f(x)$, particularly at small $x$,
because of the different anomalous dimension properties of the evolution kernels. The relative behavior between $f(x)$ and $g(x)$ is largely unchanged after QCD evolution.

The sum rules for $f(x)$ and $f_{LL}(x)$ were discussed in the Sect.~\ref{sec:rhoTMDs} 
[see Eqs.~\eqref{sumft} and \eqref{msumft}], and our values for the spin sum and tensor charge at 
the model scale are
\begin{align}
\int_0^1 \, dx \, g(x) &= 0.56, &
\int_0^1 \, dx \, h(x) &= 0.94.
\label{sumgh}
\end{align}
We therefore find that the total valence quark and antiquark contribution to the spin of the $\rho$ meson 
is $56$\%, which implies a substantial contribution of 44\% from quark orbital  angular momentum. 
Our result for the $\rho$ meson's tensor charge at the model scale is $0.94$, which is considerably 
larger than the spin sum and close to the naive quark model expectation of unity. 

Our result for the tensor polarized PDF, $f_{LL}(x)$, is symmetric around $x=1/2$ at the model scale 
(see left panel of Fig.~\ref{fig:pdfs}), being negative in the small and large $x$ regions and positive  
around the peak at $x=1/2$. Therefore, $f_{LL}(x)$ has two nodes at the model scale, which persist after 
QCD evolution, albeit continuously moving to smaller $x$ with increasing $Q^2$. The average magnitude of 
$f_{LL}(x)$ is about $0.3$ at the model scale, which is about 30\% the size of the unpolarized 
PDF.\footnote{To appreciate this point, we note that the integral of $|f_{LL}(x)|$ becomes 
$\int_0^1 dx \, |f_{LL}(x)| = 0.275$ at the model scale.} The fact that we find a tensor polarized PDF which is comparable in size to the other three PDFs, again points to the fact that there is considerable relative orbital motion of the quark and 
antiquark in the $\rho$-meson. 

After QCD evolution to $Q^2 = 5$ GeV$^{2}$ typical values for $x\lf|f_{LL}(x)\rg|$ are around $0.03$. Comparison with the HERMES Collaboration data for the $b_{1}$ structure function of the deuteron~\cite{Airapetian:2005cb}, therefore suggests that the $b_{1}$ structure function of the $\rho$ meson is roughly one order of magnitude larger than the deuteron's, likely because the nucleon mass is much larger than the
dressed quark mass. It is interesting to note that the qualitative behavior of our tensor polarized PDF $f_{LL}(x)$ is similar to the HERMES Collaboration data, and also the Bethe-Salpeter calculation for $b_1(x)$ given in Ref.~\cite{Umnikov:1996qv}, up to an overall minus sign. Therefore, one may conjecture that the $b_{1}$ structure function of the $\rho$-meson and deuteron have the opposite sign, similar to what is found for the quadrupole form factors of the $\rho$ meson~\cite{Carrillo-Serrano:2015uca,Owen:2015gva} and the deuteron~\cite{Gilman:2001yh}.

%=====================================================================
\begin{figure}[tbp]
\centering\includegraphics[width=\columnwidth]{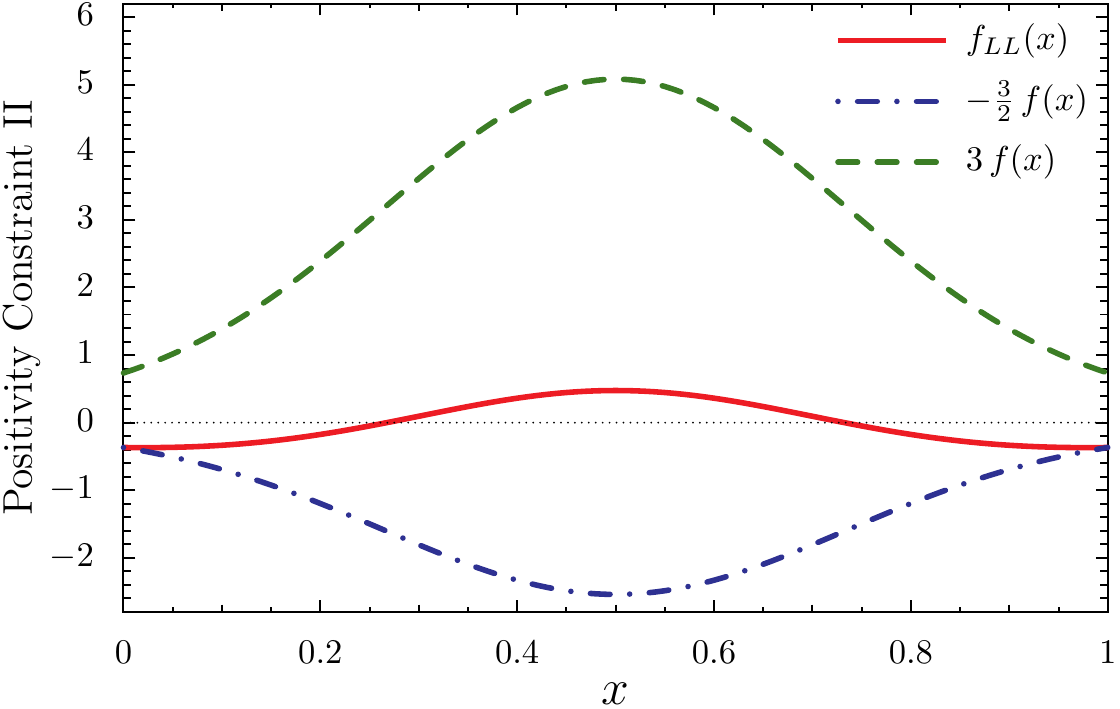}  \\[2.0ex]
\centering\includegraphics[width=\columnwidth]{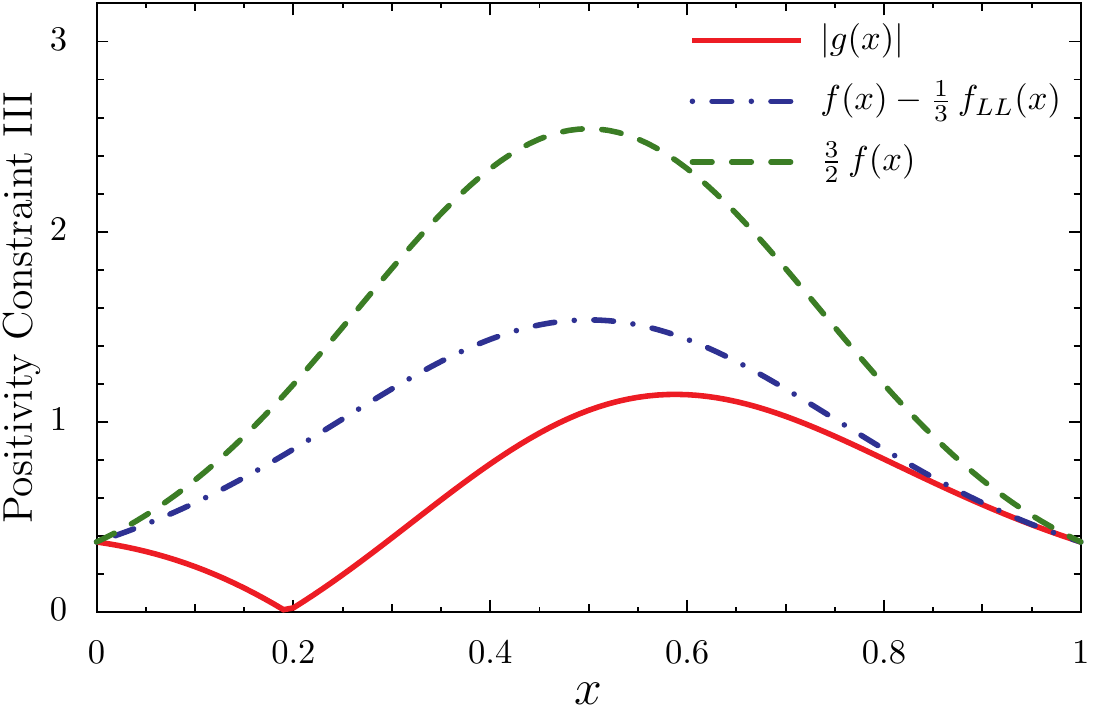} \\[2.0ex]
\centering\includegraphics[width=\columnwidth]{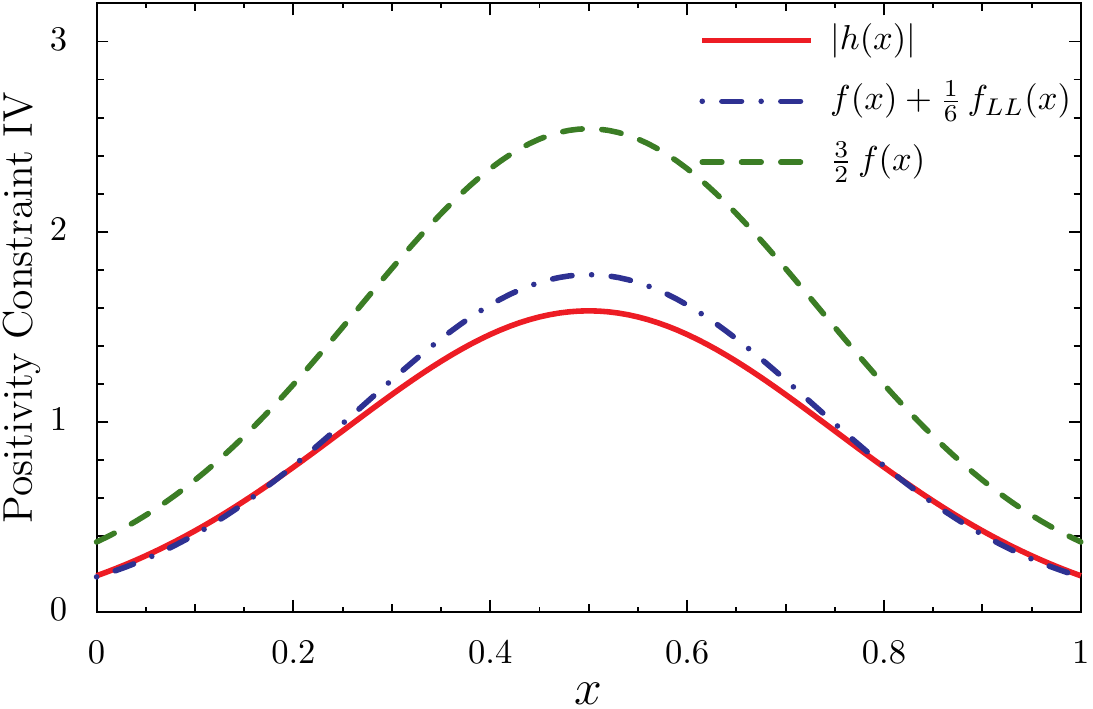}
\caption{(Color online) The positivity constraints expressed by Eq.~\eqref{bound_PDF-2} (upper panel),
Eq.~\eqref{bound_PDF-3} (middle panel) and Eq.~\eqref{bound_PDF-4} (lower panel). At the end-points of the PDFs where $x=\{0,1\}$, we find the results $f_{LL} = -\frac{3}{2}\,f$, $\lf|g\rg| = \frac{3}{2}\,f$ and $\lf|h\rg| = \frac{M}{m_\rho}\frac{3}{2}\,f$.
} 
\label{fig:positivity}
\end{figure}
%=====================================================================

In Fig.~\ref{fig:positivity} we illustrate the positivity constraints on the PDFs given by  Eqs.~\eqref{bound_PDF-1}--\eqref{bound_PDF-4}. The result in Fig.~\ref{fig:pdfs}, as well as the top and  middle panels of Fig.~\ref{fig:positivity}, demonstrate that constraints I--III are satisfied in our calculations.  The lower panel of Fig.~\ref{fig:positivity} shows positivity constraint IV for the transversity PDF,  and we see that our calculation very slightly violates this constraint in the small and large $x$ regions. However, this violation is parameter dependent, and can be remedied by choosing a slightly smaller value for the dressed quark mass, e.g. $M \simeq 375\,$MeV, which is consistent with numerous previous studies~\cite{Vogl:1991qt,Klevansky:1992qe}.

%===============================================================================
%===============================================================================
\section{SUMMARY\label{Conclusion}}
A covariant formalism for the leading-twist TMDs of spin-one hadrons has been developed, where particular emphasis was placed on developing a clear and workable definition of the TMDs which does not restrict the direction of the spin quantization axis. For the first time results for all nine $T$-even TMDs and the corresponding four PDFs of a spin-one target, in this case the $\rho^+$ meson, have been presented. The calculations were performed in the Lorentz covariant and confining framework provided by the NJL model regularized using the proper-time scheme. The importance of Lorentz covariance in satisfing the sum rules for the unpolarized and tensor polarized PDFs was demonstrated, and accordingly the baryon number and momentum sum rules for the unpolarized and tensor polarized PDFs are exactly satisfied in our calculations. The absence of unphysical thresholds for the decay of a hadron into quarks in our calculations, i.e. a manifestation of quark confinement, proved particularly important for the $\rho$-meson since its mass is very near the sum of the masses of its constituents.

Of the nine possible leading-twist $T$-even TMDs of a spin-one target, defined by Eqs.~\eqref{form1}--\eqref{form3}, the pretzelosity TMD $h_T^{\perp}(x, \vect{k}_T^2)$ vanishes identically in our $\rho^+$ calculations, and interestingly the eight remaining TMDs do not exhibit the familiar Gaussian behavior as a function of the quark transverse momentum $\vect{k}_T$. A universal feature of all non-vanishing TMDs is that as the quark transverse momentum $\vect{k}_T^2$ gets larger, the TMDs develops a much weaker dependence on the light-cone momentum fraction $x$. That is, if a quark has large $\vect{k}_T^2$ it becomes much less sensitive to $x$-dependent correlations. 

The $\rho^+$ TMDs that proved to be particularly interesting were those associated with its tensor polarization, and therefore only appear for hadrons with $J \geqslant 1$. In constrast to the TMDs that also appear for a spin-half hadron, we found that the TMD $f_{LT}(x,\vect{k}_T^2)$ vanishes identically along the $x=1/2$ line for all $\vect{k}_T^2$, and the TMD $f_{LL}(x,\vect{k}_T^2)$ vanishes along two curves in the $(x, \vect{k}_T^2)$-plane which are symmetric about the
line $x=1/2$. These are clear indications that these TMDs are very sensitive to quark orbital angular momentum. This is most easily seen by noting that $s$-wave components in a bound-state wave function usually peak at zero relative momentum, which for two equal mass constituents corresponds to $x=1/2$. 
Therefore, since these TMDs vanish near the $x=1/2$ line, but are in general non-zero, 
implies they must receive significant contributions from components of the $\rho^+$ wave function with $L \geqslant 1$. 
These findings for the tensor polarzied TMDs are a 3-dimensional generalization of earlier analysis which found that the tensor polarized structure function, $b_1(x)$, is a sensitive measure of quark orbital angular momentum~\cite{Hoodbhoy:1988am,Close:1990zw,Umnikov:1996qv}.

Integrating our TMD results over the quark transverse momentum $\vect{k}_T$ gives the four PDFs associated with a spin-one hadron. At the model scale we find that the unpolarized and transversity PDFs have a similar behavior, and that the $\rho^+$ tensor charge has the value $0.94$, which is only slightly less than the {\it naive} quark model expectation of unity. By constrast, we found that 44\% of the spin of the $\rho$ meson is carried by quark orbital angular momentum, which represents a significant departure from the {\it naive} expectation of unity. This large fraction of quark orbital angular momentum manifests as a tensor polarized PDF $f_{LL}(x)$ which is of comparable size to the three other leading-twist $\rho^+$ PDFs. In a qualitative comparison between our result for the tensor polarized $b_1$ structure function of the $\rho^+$, with the HERMES Collaboration data for the deuteron we find similar behavior, except for a relative minus sign.  
We therefore conjecture that the $b_1$ structure function of the $\rho^+$ and deuteron have opposite signs, in analogy to similar findings for the respective quadrupole moments~\cite{Carrillo-Serrano:2015uca,Owen:2015gva,Gilman:2001yh}.

For further work it would be interesting to investigate the sea quark  contributions to the tensor polarized PDF and the associated structure function, and to  estimate the deviation of the sum rule from the valence quark value given in Eq.~\eqref{sumft}. In our framework these effects could be naturally  described by taking into account the effect of the virtual pion cloud surrounding the dressed quarks.

The interesting features of the tensor polarizied TMDs found in this study, in particular the connection to quark orbital angular momentum,  warrant further investigation. A key avenue is the experimental and theoretical study of the deuteron tensor polarized TMDs, where we already have hints from the HERMES $b_1$ structure function data that the small $x$ region may be particularly interesting~\cite{Miller:1989nc,Liuti:2014dda,Miller:2013hla}. In addition, the TMD FFs for the production of vector mesons provides another interesting possibility to study the 3-dimensional tensor structure of hadrons. An ideal facility for such measurements would be the proposed electron-ion collider.

\begin{acknowledgments}
This work was supported by the U.S. Department of Energy, Office of Science, Office of Nuclear Physics, contract no. DE-AC02-06CH11357; and Laboratory Directed Research and Development (LDRD) funding from Argonne National Laboratory, project no. 2016-098-N0 and project no. 2017-058-N0. Y. N. wishes to express his thankfulness to his parents for the great support of his Ph.D. course. Y.N. also appreciates Tokai University for a Ph.D. course scholarship.
\end{acknowledgments}

%===============================================================================
%===============================================================================
\appendix
\section{POSITIVITY CONSTRAINTS FOR SPIN-ONE TMDS\label{app:relations}}
Here we follow the methods of Refs.~\cite{Bacchetta:2000jk,Bacchetta:2001rb} to derive the positivity constraints on the
spin-one TMDs. From the operator definition given in Eq.~\eqref{phi1} it follows that
\begin{align}
\lf[ \Phi^{(\l)_{\vect{S}}}(x, \vect{k}_T) \g^+ \rg]_{\beta \alpha}
&= \sqrt{2} \int \frac{dz^- \, d^2 \vect{z}_T}{(2\pi)^3} \,  
e^{i\,xp^+\,z^-} e^{-i\,\vect{k}_T \cdot\, \vect{z}_T} \no \\
&\hs*{-8mm}
\times {}_{\vect{S}}\langle p,\l | {\psi}^{\dagger}_{(+) \alpha}(0) \,\psi_{(+) \beta}(z^-,\vect{z}_T) | p, \l \rangle_{\vect{S}},
\label{phi4}
\end{align}
where $\psi_{(+)} = \L_{(+)} \psi$ (with $\L_{(+)}=\g^0 \g^+/\sqrt{2}$) 
are the good components of the quark field operator~\cite{Jaffe:1991ra}.
We can still represent this quantity by Fig. \ref{quark-correlation-function} if the 
external quark lines refer to the good components. Viewed in   
this way, the matrix $\lf[ \Phi^{(\l)_{\vect{S}}}(x, \vect{k}_T) \g^+ \rg]_{\beta \alpha}$,
is the {\em transpose} of the hadron-antiquark forward scattering matrix (we call it
$M$), apart from overall normalization:
\begin{align}
M^{(\l)_{\vect{S}}} (x, \vect{k}_T) = \lf[ \Phi^{(\l)_{\vect{S}}}(x, \vect{k}_T) \g^+ \rg]^T .
\label{m1}
\end{align} 
The positivity conditions must be imposed on this matrix $M$.

At leading-twist the matrix ${\Phi}\,\g^+$ takes the form [see Eqs.~\eqref{exp}--\eqref{tp}] 
\begin{align}
&\Phi^{(\l)_{\vect{S}}} (x, \vect{k}_T)\, \g^+ = \no \\
&
\lf[
\langle \g^+ \rangle^{(\l)}_{\vect{S}} 
+ \langle \g^+ \g_5 \rangle^{(\l)}_{\vect{S}} \cdot \g_5 
- \langle \g^+ \vect{\g}_T \g_5 \rangle^{(\l)}_{\vect{S}} \cdot \g_5 \vect{\g}_T  \rg] \L_{(+)}.
\label{phi3}
\end{align}
It can be reduced to a $2 \times 2$ matrix by using the chiral representation of
the Dirac matrices. By making an orthogonal transformation so that only
the upper left $2 \times 2$ block is nonzero, and taking the
transpose according to Eq.~\eqref{m1}, we arrive at  
\begin{align}
M^{(\l)_{\vect{S}}} (x, \vect{k}_T)
= 
\begin{bmatrix}
\langle \g^+ \rangle^{(\l)}_{\vect{S}} +  \langle \g^+ \g_5 \rangle^{(\l)}_{\vect{S}} & 
\langle \g^+ \, \g_T^{(+)} \g_5 \rangle^{(\l)}_{\vect{S}}  \\[0.5em]
\langle \g^+ \, \g_T^{(-)} \g_5 \rangle^{(\l)}_{\vect{S}} &  
\langle \g^+ \rangle^{(\l)}_{\vect{S}} -  \langle \g^+ \g_5 \rangle^{(\l)}_{\vect{S}}       
\end{bmatrix}.
\label{m2}
\end{align}
Here the various matrix elements are given in terms of the TMDs by Eqs.~\eqref{form1}--\eqref{form3}
and we have defined
\begin{align}
\langle \g^+ \g_T^{(\pm)} \g_5  \rangle^{(\l)}_{\vect{S}} =  \langle \g^+ \g^{1} 
\g_5  \rangle^{(\l)}_{\vect{S}}
\pm i \langle \g^+ \g^{2} \g_5  \rangle^{(\l)}_{\vect{S}}  .
\label{tpm}
\end{align}
In the rest frame of the hadron, each of the four matrix elements in Eq.~\eqref{m2} is of the form 
\begin{align}
M^{(\l)_{\vect{S}}}_{\alpha \beta} = \vect{\ve}^{\dagger j}_{(\l)} \, M_{\alpha \beta}^{ji} \, \vect{\ve}_{(\l)}^{i} 
= {\rm Tr} \lf(M_{\alpha \beta} \, \rho_{(\l)} \rg),
\label{trace2}
\end{align}
where the polarization 3-vectors $\vect{\ve}_{(\l)}$ refer to an arbitrary
direction $\vect{S}$ as explained in Sect.~\ref{form-Nino}, and the matrix $M_{\alpha \beta}^{ji}$
is independent of $\l$ and $\vect{S}$.  
In Eq.~\eqref{trace2} we introduced the spin density matrix for a pure spin state of the hadron in the rest frame, 
with spin projection $\l$ on the direction $\vect{S}$, according to [see Eq.~\eqref{formula}]
\begin{align}
\rho_{(\l)}^{ij} &= \vect{\ve}_{(\l)}^i \,\vect{\ve}_{(\l)}^{\dagger j}
= \frac{1}{3} \lf( 1 + \frac{3}{2} \l \lf(\vect{\Sigma} \cdot \vect{S} \rg)
\rg)^{ij} - T^{ij}_{(\l)},
\label{spind} 
\end{align}
where the symmetric and traceless tensor $T^{ij}_{(\l)}$ is defined by
\begin{align}
T^{ij}_{(\l)} = \frac{3 \l^2 -2}{2} \lf(S^i S^j - \frac{1}{3} \delta^{ij}\rg)\,.
\label{sd2}
\end{align}
Following Ref.~\cite{Bacchetta:2000jk}, we introduce the matrices
\begin{align}
\lf(\Sigma^{kl}\rg)^{ij} = \lf[ \frac{1}{2} \lf( \Sigma^k \Sigma ^l
+ \Sigma^l \Sigma^k \rg) - \frac{2}{3} \delta^{kl} \ident\rg]^{ij}.
\label{skl}
\end{align}
The spin density matrix Eq.~\eqref{spind} then can be expressed by\footnote{This is the analogue of Eq.~(8) in Ref.~\cite{Bacchetta:2001rb} for a pure spin state of the hadron characterized by the spin projection 
$\lambda$ onto the arbitrarily chosen direction $\vect{S}$.
In fact, if we parametrize the tensor of Eq.~\eqref{sd2} in the form of Eq.~(3) of Ref.~\cite{Bacchetta:2001rb} as
\begin{align}
T_{(\lambda)}^{ij} = \frac{1}{2}
\begin{pmatrix}
S_{LL} + S_{TT}^{11} &  S_{TT}^{12}  &  S_{LT}^1  \\[0.3em]
S_{TT}^{12}  &  S_{LL} + S_{TT}^{22}  &  S_{LT}^2  \\[0.3em]
S_{LT}^1  &  S_{LT}^2  &  -2 S_{LL}   
\end{pmatrix},
\end{align}
with $S_{TT}^{22} = - S_{TT}^{11}$, then the quantities $S_{LL}$, $S_{LT}^i$ and $S_{TT}^{ij}$ 
precisely agree with the forms given in Eqs.~\eqref{sll}--\eqref{stt}.} 
\begin{align}
\rho^{ij}_{(\l)} = \frac{1}{3} \lf[ 1 + \frac{3}{2} \l
\lf(\vect{\Sigma} \cdot \vect{S}\rg) 
+ 3\lf(T_{(\l)}^{kl}\,\Sigma^{kl}\rg) \rg]^{ij}.
\label{sd3}
\end{align}
It is easy to check that the spin density matrix of Eq.~\eqref{sd3}
satisfies the relations
\begin{align}
{\rm Tr} \, \rho_{(\l)} &= 1,  \label{p1} \\
{\rm Tr} \lf( \rho_{(\l)} \, \Sigma^k \rg) &= \l\,S^k,  
\label{p2} \\
{\rm Tr} \lf( \rho_{(\l)} \, \Sigma^{kl}\rg) &= T_{(\l)}^{kl}.
\label{p3} 
\end{align}
Returning now to the problem of constructing the matrices $M_{\alpha \beta}^{ij}$ from 
Eq.~\eqref{trace2}, we see from the relations Eqs.~\eqref{p1}--\eqref{p3} that we simply 
have to perform the replacements
\begin{align}
\l\,\vect{S} \rightarrow \vect{\Sigma} \qquad \text{and} \qquad
T_{(\l)}^{ij} \rightarrow \Sigma^{ij},
\label{rep2}
\end{align}
everywhere in Eq.~\eqref{m2}, where $\vect{\Sigma}$ and ${\Sigma}^{ij}$ act 
in the hadron spin space. We write the required matrix $M$ in the form    
\begin{align}
M (x, \vect{k}_T) = \begin{pmatrix} A & B \\ C & D  \end{pmatrix},
\label{mh2}
\end{align}
where $A, B, C, D$ are $3 \times 3$ matrices in the hadron spin space.
To derive their explicit forms, we use the regular representation
of the matrices $\vect{\Sigma}$ and $\vect{\Sigma}^{kl}$ 
to obtain finally 
\begin{align}
A &=
\begin{pmatrix}
f - \frac{1}{3}\,f_{LL} + g_L &  \frac{|\vect{k}_T|}{\sqrt{2}\,m_h}\, e^{-i \phi} \, g_T^{(+)}
&  \frac{\vect{k}_T^2}{m_h^2}\, e^{-2i\phi} \, f_{TT}    \\
\frac{|\vect{k}_T|}{\sqrt{2}\,m_h}\, e^{i \phi} \, g_T^{(+)} & f + \frac{2}{3}\, f_{LL}  &
\frac{|\vect{k}_T|}{\sqrt{2}\,m_h}\, e^{-i \phi} \, g_T^{(-)}  \\
 \frac{\vect{k}_T^2}{m_h^2}\, e^{2i\phi} \, f_{TT} 
& \frac{|\vect{k}_T|}{\sqrt{2}\,m_h}\, e^{i \phi} \, g_T^{(-)} & f - \frac{1}{3}\, f_{LL} - g_L      
\end{pmatrix},  \\
B &= 
\begin{pmatrix}
\frac{|\vect{k}_T|}{m_h}\,  e^{i\phi} \, h_L^{\perp}  &  \sqrt{2} \, h  &   0  \\
\frac{\vect{k}_T^2}{\sqrt{2}\,m_h^2}\, e^{2i\phi} \, h_T^{\perp}  &  0  &  \sqrt{2} \, h  \\
0  &  \frac{\vect{k}_T^2}{\sqrt{2}\,m_h^2}\, e^{2i\phi} \, h_T^{\perp}  &  
- \frac{|\vect{k}_T|}{m_h}\,  e^{i\phi} \, h_L^{\perp}   
\end{pmatrix}, \\[0.7em]
C &= 
\begin{pmatrix}
\frac{|\vect{k}_T|}{m_h} \, e^{-i\phi} \, h_L^{\perp}  &  
\frac{\vect{k}_T^2}{\sqrt{2}\,m_h^2}\, e^{-2i\phi} \, h_T^{\perp} & 0  \\ 
\sqrt{2} \, h  &   0 &  \frac{\vect{k}_T^2}{\sqrt{2}\,m_h^2}\, e^{-2i\phi} \, h_T^{\perp}  \\
0  &  \sqrt{2} \, h  &  - \frac{|\vect{k}_T|}{m_h}\,  e^{-i\phi} \, h_L^{\perp}           
\end{pmatrix}, \\[0.7em]
D &= 
\begin{pmatrix}
f - \frac{1}{3}\, f_{LL} - g_L &  -\frac{|\vect{k}_T|}{\sqrt{2}\, m_h}\, e^{-i \phi} \, g_T^{(-)}
&  \frac{\vect{k}_T^2}{m_h^2}\, e^{-2i\phi} \, f_{TT}    \\
- \frac{|\vect{k}_T|}{\sqrt{2}\, m_h}\, e^{i \phi} \, g_T^{(-)} & f + \frac{2}{3}\, f_{LL}  &
- \frac{|\vect{k}_T|}{\sqrt{2}\, m_h}\, e^{-i \phi} \, g_T^{(+)}  \\
 \frac{\vect{k}_T^2}{m_h^2}\, e^{2i\phi} \, f_{TT} 
& - \frac{|\vect{k}_T|}{\sqrt{2}\, m_h}\, e^{i \phi} \, g_T^{(+)} & f - \frac{1}{3}\, f_{LL} + g_L      
\end{pmatrix},
\end{align}
where
\begin{align}
g_T^{(\pm)} = g_T \pm  f_{LT},
\end{align}
and $\phi$ is the azimuthal angle defined by $k_T = \lf|\vect{k}_T\rg|\lf(\cos \phi, \sin \phi\rg)$.
By construction, $A$ and $D$ are hermite, and $C = B^{\dagger}$.
The positivity conditions for the TMDs are then equivalent to the requirement
that all six eigenvalues of the matrix given by Eq.~\eqref{mh2} are semi-positive for all values
of $x$ and $\vect{k}_T^2$.\footnote{See Ref.~\cite{Bourrely:1980mr} for a general discussion of positivity conditions.}
Therefore, the nine $T$-even TMDs for a spin-one target should satisfy 6 (sufficient) positivity conditions, however from the positivity of the principal minors of the matrix given in Eq.~\eqref{mh2} it is possible to obtain numerous necessary conditions. For example, positivity of the 1-dimensional principal minors gives
\begingroup
\allowdisplaybreaks
\begin{align}
\label{eq:tmdpos1}
f + \frac{2}{3}\,f_{LL} &\geqslant 0, \\
f + g_L - \frac{1}{3}\,f_{LL} &\geqslant 0, \\
 f - g_L - \frac{1}{3}\,f_{LL} &\geqslant 0,
\end{align}
\endgroup
and positivity of the 2-dimensional principal minors implies:
\begingroup
\allowdisplaybreaks
\begin{align}
\label{eq:tmdpos4}
\lf(f + \frac{2}{3}\,f_{LL}\rg) \lf(f + g_L - \frac{1}{3}\,f_{LL}\rg) - 2\,h^2 &\geqslant 0, \\
\lf(f - \frac{1}{3}\,f_{LL}\rg)^2 - g_L^2 - 4 f_{TT}^2\,\hat{k}_m^4 &\geqslant 0, \\
\lf(f - \frac{1}{3}\,f_{LL}\rg)^2 - g_L^2 - 2\, h_L^\perp{}^2\, \hat{k}_m^2 &\geqslant 0, \\
\lf(f + \frac{2}{3}\,f_{LL}\rg) \lf(f - g_L- \frac{1}{3}\,f_{LL}\rg) - 2\, h_T^\perp{}^2\, \hat{k}_m^4 &\geqslant 0, \\
\lf(f + \frac{2}{3}\,f_{LL}\rg) \lf(f - g_L- \frac{1}{3}\,f_{LL}\rg) - g_T^{(-)}{}^2\,\hat{k}_m^2 &\geqslant 0, \\
\lf(f + \frac{2}{3}\,f_{LL}\rg) \lf(f + g_L- \frac{1}{3}\,f_{LL}\rg) - g_T^{(+)}{}^2\, \hat{k}_m^2 &\geqslant 0,
\end{align}
\endgroup
where $\hat{k}_m \equiv \frac{\lf|\vect{k}_T\rg|}{\sqrt{2}\,m_h}$ and we have dropped the $(x,\vect{k}_T^2)$-dependence of the TMDs to aid clarity. Note, Eqs.~\eqref{eq:tmdpos1}--\eqref{eq:tmdpos4} imply the naive positivity conditions given in Eqs.~\eqref{eq:naive1}--\eqref{eq:naive4}.

For the $\vect{k}_T$-integrated case the matrices $A$, $B$, $C$ and $D$ take the simple form
\begingroup
\allowdisplaybreaks
\begin{align}
A(x) &= 
\begin{pmatrix}
\alpha (x)  &  0        &  0  \\
0           &  \beta(x) &  0  \\
0           &  0        &  \g(x)          
\end{pmatrix}, \\[0.5em]
B(x) &=
\begin{pmatrix}
0  &  \sqrt{2}\,h(x)  &  0  \\
0  &  0               &  \sqrt{2}\,h(x)  \\
0  &  0               &  0           
\end{pmatrix},  \\[0.5em]
C(x) &=
\begin{pmatrix}
0               &  0               &  0  \\
\sqrt{2}\,h(x)  &  0               &  0  \\
0               &  \sqrt{2}\,h(x)  &  0           
\end{pmatrix}, \\[0.5em]
D(x) &= 
\begin{pmatrix}
\g(x) &  0        &  0  \\
0     &  \beta(x) &  0  \\
0     &  0        &  \alpha(x)          
\end{pmatrix},
\end{align}
\endgroup
where we have defined 
\begin{align}
\alpha (x) &\equiv f(x) - \frac{1}{3}\,f_{LL}(x) + g(x), \\
\beta (x) &\equiv f(x) + \frac{2}{3}\,f_{LL}(x), \\
\g(x) &\equiv f(x) - \frac{1}{3}\,f_{LL}(x) - g(x).
\end{align}
The positivity of the eigenvalues of the matrix given in Eq.~\eqref{mh2}, for the $\vect{k}_T$-integrated case, gives three {\it sufficient} positivity conditions for the PDFs of spin-one hadrons. These conditions can be expressed as
\begingroup
\allowdisplaybreaks
\begin{align}
\label{eq:pos1}
&f(x) - g(x) - \frac{1}{3}\,f_{LL}(x) \geqslant 0,  \\
\label{eq:pos2}
&2\,f(x) + g(x) + \frac{1}{3}\,f_{LL}(x) \geqslant 0, \\
\label{soff1} 
&\lf(f(x) + \frac{2}{3}\,f_{LL}(x)\rg)\lf(f(x) + g(x) - \frac{1}{3}\,f_{LL}(x)\rg) \geqslant 2\,h(x)^2,
\end{align}
\endgroup
where the final inequality is the generalization of the Soffer bound~\cite{Soffer:1994ww} for a spin-one target~\cite{Bacchetta:2000jk,Bacchetta:2001rb}. By noting that the two expressions in the brackets of Eq.~\eqref{soff1} must be non-negative, otherwise there is a contradiction with Eq.~\eqref{eq:pos2}, it is straightforward to obtain the {\it naive} positivity conditions given in Eqs.~\eqref{bound_PDF-1}--\eqref{bound_PDF-3}. Then, if we use the relation $g(x) \leqslant f(x) - \frac{1}{3} f_{LL}(x)$ in the second expression on the left-hand side of Eq.~\eqref{soff1}, and the result $f(x) + \frac{1}{6}\,f_{LL}(x) \geqslant 0$, the condition given by Eq.~\eqref{soff1} leads to
\begin{align}
|h(x)| \leqslant f(x) + \frac{1}{6}\,f_{LL}(x) \leqslant \frac{3}{2}\,f(x),
\end{align}
which is the {\it naive} positivity relation given in Eq.~\eqref{bound_PDF-4}.

%===============================================================================
%===============================================================================
\section{EXPLICIT RESULTS FOR RHO TMDS AND PDFS\label{ex-form}}
For the explicit calculation of the leading-twist TMDs of the spin-one $\rho^+$ meson we have used the NJL model regularized using the 
proper-time scheme. In this regularization scheme the denominators of the loop integrals are replaced by
\begin{align}
\frac{1}{D^{n}} &= \frac{1}{(n-1)!} \int^{\infty}_{0} d\tau\,\tau^{n-1}\,e^{-\tau D} \\
&
\rightarrow  \frac{1}{(n-1)!} 
\int^{1/\L_{IR}^2}_{1/\L_{UV}^2} d\tau \,\tau^{n-1}\, e^{-\tau D},
\label{proper}
\end{align}
where $\L_{UV}$ is the ultraviolet and $\L_{IR}$ the infrared cut-off.
The gap equation given in Eq.~\eqref{gap} then takes the form
\begin{align}
M = m + \frac{3 G_{\pi}}{\pi^2} \, \int d \tau\ \frac{e^{-\tau M^2}}{\tau^2},  
\label{gap1}
\end{align}
where here, and in the following, we drop the regularization parameters on the proper-time integration to aid clarity. The bubble graph of Eq.~\eqref{bub} in the proper-time regularization scheme becomes:
\begin{align}
\Pi_{\rho}(p^2) = -\frac{3 p^2}{\pi^2} \int_0^1 dx \int
\frac{d \tau}{\tau} \
x(1-x) \, e^{-\tau \lf[M^2 - x(1-x)p^2\rg]}.
\label{piapp}
\end{align}
The form of the coupling constant $Z_\rho$ can then be obtained from Eq.~\eqref{grho} as
\begin{align}
Z_\rho^{-2} &=  \frac{3}{\pi^2} \int_0^1 dx \int d \tau \no \\
&\hs*{0mm}
\times \
 x(1-x)\lf[\frac{1}{\tau} + x(1-x) m_{\rho}^2 \rg]
 e^{-\tau \lf[M^2 -  x(1-x)\,m_{\rho}^2\rg]}.
\label{gapp}
\end{align}
We now present the explicit results for the $\rho^+$ meson TMDs and PDFs. Firstly, we give results for the Lorentz tensors $\lf<\G \rg>^{\mu\nu}\!(x,\vect{k}_T)$ of Eq.~\eqref{NJL-form}. For $\G = \g^+$ we obtain\footnote{Note, there is no summation over $i$ in the term proportional to $f_{5a}$ in Eq.~\eqref{gpapp}.}
\begin{align}
&\hs{-5mm}\lf<\g^{+} \rg>^{\mu\nu} (x,\vect{k}_T) \equiv g^{\mu\nu}\, f_1(x,\vect{k}_T^2)+ 
\frac{g^{\mu+}g^{\nu+}}{p^+p^+} \,p^2\, f_2(x,\vect{k}_T^2) \no \\
& + \frac{g^{\mu+}g^{\nu-}+g^{\mu-}g^{\nu+}}{p^+}\ f_3(x,\vect{k}_T^2) \no \\
&
+ \frac{g^{+\mu}g^{\nu i}+ g^{+\nu}g^{\mu i}}{p^+}\, k_{Ti}\, f_4(x,\vect{k}_T^2)  \no \\
&
+ g^{\mu i}g^{\nu i}f_{5a}(x,\vect{k}_T^2)
- g^{\mu i}g^{\nu j}\,k_{Ti}\,k_{Tj}\,f_{5b}(x,\vect{k}_T^2)  \no\\
& +\frac{p^{\mu}p^{\nu}}{p^2}\, f_6(x,\vect{k}_T^2) + \frac{g^{\mu+}p^{\nu}+ g^{\nu+}p^{\mu}}
{p^+}\, f_7(x,\vect{k}_T^2) \no \\
&+ \lf(g^{\mu i}p^{\nu} + g^{\nu i}p^{\mu}\rg) k_{Ti}\, f_8(x,\vect{k}_T^2),
\label{gpapp}
\end{align}
where we defined 
\begingroup
\allowdisplaybreaks
\begin{align}
f_{1}(x,\vect{k}_T^2) &= -\frac{3\,Z_\rho^2}{4\,\pi^3} 
\int d\tau \ e^{-\tau\,A} \no \\
&\hs*{12mm} \times
\biggl[1 - 2\,x(1-x) + x(1-x)p^2 \tau \biggr], \\
f_{2}(x,\vect{k}_T^2) &= -\frac{3\,Z_\rho^2}{4\,\pi^3} \int  
d\tau \ e^{-\tau\,A} \no \allowdisplaybreaks[0]\\
&\hs*{-1mm}
\times \lf[1 - 6x(1-x) + x(1-x)(1-2x)^2 p^2\tau\rg], \\
f_{3} (x,\vect{k}_T^2) &= 0, \\
f_{4}(x,\vect{k}_T^2) &= -\frac{3\,Z_\rho^2}{4\,\pi^3} \int  
d\tau\ e^{-\tau\,A} \no \\
&\hs*{15mm}
\times (1-2x)\lf[1 + 2\,x(1-x)p^2 \tau\rg], \\
f_{5a}(x,\vect{k}_T^2) &= \frac{3\,Z_\rho^2}{2\,\pi^3} \int  
d\tau\ e^{-\tau\,A}\,
x(1-x), \\
f_{5b}(x,\vect{k}_T^2) &= \frac{3\,Z_\rho^2}{\pi^3} \int  
d\tau\ e^{-\tau\,A}\,
x(1-x)\,\tau, \\
f_{6}(x,\vect{k}_T^2) &= \frac{3\,Z_\rho^2}{\pi^3} \int  
d\tau e^{-\tau\,A}\,x^2(1-x)^2 p^2, \\
f_{7}(x,\vect{k}_T^2) &= \frac{3\,Z_\rho^2}{4\,\pi^3} \int  
d\tau \ e^{-\tau\,A} \no \\
&\hs*{-15mm}
\times \lf[2\,x(1-x) + \lf[1-6x(1-x)+x(1-x)(1-2x)^2 p^2\tau\rg]\rg], \\
f_8(x,\vect{k}_T^2) &= \frac{3\,Z_\rho^2}{2\,\pi^3} \int d\tau\ 
e^{-\tau\,A}\, x(1-x)(1-2x)\,\tau.
\end{align}
\endgroup
For $\G = \g^+ \g_5$ we obtain
\begin{align}
&\lf<\g^{+}\g_{5} \rg> ^{\mu\nu} (x, \vect{k}_T) \equiv i\ve^{+\mu\nu-}\,g_1(x,\vect{k}_T^2) \no \\
&\hs*{7mm}
+ \frac{i\ve^{+\mu\nu i} }{p^+}\,k^i_T\, g_2(x,\vect{k}_T^2) 
+ i\ve^{\a\mu\nu i}\,p_\a\,k^i_T\,g_3(x, \vect{k}_T^2),
\end{align}
where 
\begin{align}
g_{1}(x, \vect{k}_T^2) &= \frac{3\,Z_\rho^2}{4\,\pi^3} \int  
d\tau\ e^{-\tau\,A} \no\\
&\hs*{7mm}
\lf[1 + x(1-x)p^2\tau - 2(1-x)\,\vect{k}_T^2\tau\rg], \\
g_{2}(x, \vect{k}_T^2) &=  -\frac{3\,Z_\rho^2}{4\,\pi^3} \int  
d\tau\ e^{-\tau\,A}, \\
g_{3}(x, \vect{k}_T^2) &=  -\frac{3\,Z_\rho^2}{2\,\pi^3} \int  
d\tau\ e^{-\tau\,A} \ x(1-x)\,\tau.
\end{align}
For $\G = \g^+ \g^i \g_5$ we obtain
\begin{align}
&\lf<\g^{+}\g^{i}\g_5 \rg> ^{\mu\nu}\!(x,\vect{k}_T) \equiv 
\frac{1}{p^+}\,i\ve^{\mu\nu i +}\,h_1(x,\vect{k}_T^2)\hs*{25mm} \no \\
&\hs*{14mm}
+ \frac{1}{p^+}\lf[g^{\mu +}i\ve^{\nu i-+}-g^{\mu +}i\ve^{\nu i-+}\rg] h_2(x,\vect{k}_T^2) \no\\
&\hs*{14mm}
+ \lf[g^{\mu j}\,k_{Tj}\, i\ve^{\nu i-+} - g^{\nu j}\,k_{Tj}\,i\ve^{\mu i-+}\rg]h_3(x, \vect{k}_T^2) \no \\
&\hs*{14mm}
+ \lf[p^{\mu} i\ve^{\nu i-+} - p^{\nu }i\ve^{\mu i-+}\rg] h_4(x, \vect{k}_T^2),
\end{align}
where 
\begingroup
\allowdisplaybreaks
\begin{align}
h_{1}(x, \vect{k}_T^2) &= \frac{3\,M\,Z_\rho^2}{4\,\pi^3} \int  
d\tau\ e^{-\tau\,A}\,(1-x)p^2\tau, \\
h_{2}(x, \vect{k}_T^2) &= \frac{3\,M\,Z_\rho^2}{4\,\pi^3} \int  
d\tau\ e^{-\tau\,A}\lf[1 - (1-x)(1-2x)p^2\tau\rg], \\
h_{3}(x, \vect{k}_T^2) &= -\frac{3\,M\,Z_\rho^2}{2\,\pi^3} \int  
d\tau\ e^{-\tau\,A}\,(1-x)\,\tau, \\
h_{4}(x, \vect{k}_T^2) &= -\frac{3\,M\,Z_\rho^2}{2\,\pi^3} \int  
d\tau\ e^{-\tau\,A}\, x(1-x)\,\tau.
\end{align}
\endgroup
In all the expressions given above $A = \vect{k}_T^2 +M^2-x(1-x)p^2$ and these results hold completely off-shell. By integrating Eq.~\eqref{gpapp} over $x$ and $\vect{k}_T$, it is easy to confirm the result of the Ward identity given in Eq.~\eqref{ward}.

The $\rho^+$ meson TMDs can be obtained by contracting these formulas with Eq.~\eqref{formula}. 
The resulting TMDs are
\begingroup
\allowdisplaybreaks
\begin{align}
\label{eq:f1}
f(x,\vect{k}_T^2) &= \frac{Z_\rho^2}{2\,\pi^3} \int  
d\tau\ e^{-\tau\,A}\,\Bigl[2\,x(1-x)\lf(1 - \tau\,\vect{k}_T^2\rg) \no \\
&\hs*{3.5mm}
+ 1 + x(1-x)\lf[1 + 2\,x(1-x)\rg]m_\rho^2\,\tau\Bigr],\\
f_{LL}(x,\vect{k}_T^2) &=-\frac{3\,Z_\rho^2}{4\pi^3} \int  
d\tau\ e^{-\tau\,A} \no\\
&\hs*{-9mm}
\times \lf[1 + x(1-x)\lf[(1-2x)^2 m_\rho^2\tau - 2\,\tau\,\vect{k}_T^2 - 4\rg] \rg], \\
f_{LT}(x,\vect{k}_T^2) &= - \frac{3 Z_\rho^2}{4\,\pi^3} \int  
d\tau\ e^{-\tau\,A}\ (1-2x)\no \allowdisplaybreaks[0]\\
&\hs{28mm}
\lf[1+2x(1-x)m_\rho^2 \tau\rg], \\
f_{TT}(x,\vect{k}_T^2) &= \frac{3\,Z_\rho^2}{2 \pi^3} \int  
d\tau\ e^{-\tau\,A} \ x(1-x)\,m_\rho^2\,\tau, \\
\label{eq:gL1}
g_L(x,\vect{k}_T^2) &= \frac{3\,Z_\rho^2}{4\,\pi^3} \int  
d\tau\ e^{-\tau\,A} \no\\
&\hs*{6mm}
\lf[1 + x(1-x)\,m_\rho^2\tau - 2(1-x)\,\tau\,\vect{k}_T^2\rg], \\[0.4em]
\label{eq:gT1}
g_T(x,\vect{k}_T^2) & =\frac{3\,Z_\rho^2}{4\,\pi^3} \int  
d\tau\ e^{-\tau\,A}
\lf[1 + 2\,x(1-x)\,m_\rho^2\,\tau\rg], \\
\label{eq:h1}
h(x,\vect{k}_T^2) & =\frac{3\,Z_\rho^2\,M}{4\,\pi^3\,m_{\rho}} \int  
d\tau\ e^{-\tau\,A}\lf[1 + 2\,x(1-x)\,m_\rho^2\,\tau\rg], \\
\label{eq:hL}
h_{L}^{\perp}(x,\vect{k}_T^2) &= -\frac{3\,Z_\rho^2\,M}{2\,\pi^3\,m_{\rho}} \int  
d\tau\ e^{-\tau\,A}\,(1-x)\,m_\rho^2\,\tau, \\
h_T^{\perp}(x,\vect{k}_T^2) &= 0,
\end{align}
\endgroup
where $A = \vect{k}_T^2 +M^2-x(1-x)m_\rho^2$. Integrating these expressions over $\vect{k}_T$, we get 
the formulas for the $\rho^+$ meson PDFs as follows:
\begingroup
\allowdisplaybreaks
\begin{align}
f(x) &= \frac{Z_\rho^2}{2\,\pi^2} \int d\tau\ e^{-\tau\,B} \no \\
&\hs*{8mm}
\frac{1}{\tau}\lf[1 + x(1-x)\lf[1+2x(1-x)\rg]m_\rho^2\,\tau\rg], \label{fapp}  \\
f_{LL}(x) &= -\frac{3 Z_\rho^2}{4\pi^2} \int  d\tau\ e^{-\tau\,B}  
\no \allowdisplaybreaks[0]\\
&\hs*{0mm}
\frac{1}{\tau}  \lf [ 1-6x(1-x)+x(1-x)(1-2x)^2 m_\rho^2\tau \rg ],  \label{thetaapp} \\
g(x) &=\frac{3\,Z_\rho^2}{4\,\pi^2} \int  d\tau\ e^{-\tau\,B}
\lf[\frac{2x-1}{\tau} + x(1-x)\,m_\rho^2\rg], \\
h(x) &= \frac{3\,Z_\rho^2\,M}{4\,\pi^2\,m_{\rho}} \int  d\tau\, e^{-\tau\,B}
\lf[\frac{1}{\tau} + 2\,x(1-x)\,m_\rho^2\rg],
\end{align}
\endgroup
where $B = M^2 - x(1-x)m_\rho^2$. From Eqs.~\eqref{gapp},~\eqref{fapp} and \eqref{thetaapp}, it is easy to confirm the sum rules
given in Eqs.~\eqref{sumft} and \eqref{msumft}.
 
Finally, we remark that it is straightforward to derive the expressions for the elementary $u \rightarrow \rho^+$ fragmentation process from the formulas given in this Appendix as follows: If we compare the operator definition of Eq.~\eqref{phi1} to the corresponding definition for the fragmentation functions, as given for example by Eq.~(II.1) of Ref.~\cite{Bentz:2016rav}, we see that one can obtain the  fragmentation matrix from the TMD matrix given by Eq.~\eqref{exp} by the following 3 steps: First, replace $x \rightarrow 1/z$ , where $z$ denotes the scaling variable  for the fragmentation functions.  Second, replace $\vect{k}_T \rightarrow  - \vect{p}_{\perp}/z$, where $\vect{p}_{\perp}$ denotes the transverse momentum of the produced hadron relative to the direction of the momentum of the fragmenting quark. Third, multiply an overall factor $1/(6\,z)$.

As emphasized in Ref.~\cite{Ito:2009zc}, this prescription can be applied to
obtain the fragmentation functions for the elementary process from the
corresponding distribution functions, but in order to describe the behavior
of the empirical fragmentation functions one needs to take into account
multi-fragmentation processes~\cite{Bentz:2016rav}.

%===============================================================================
%===============================================================================
%\bibliographystyle{/home/icloet/.files/myapsrev4-1}
%\bibliography{bibtexfile,bibtexfile_cloet,bibtexfile_arXiv,bibtexfile_misc,bibtexfile_books}


\end{document}